\def\year{2019}\relax
\documentclass[letterpaper]{article} % DO NOT CHANGE THIS
\usepackage{aaai19}  % DO NOT CHANGE THIS
\usepackage{times}  % DO NOT CHANGE THIS
\usepackage{helvet} % DO NOT CHANGE THIS
\usepackage{courier}  % DO NOT CHANGE THIS
\usepackage[hyphens]{url}  % DO NOT CHANGE THIS
\usepackage{graphicx} % DO NOT CHANGE THIS
\usepackage{graphicx}  % DO NOT CHANGE THIS
\usepackage{subcaption}
\usepackage{eqnarray,amsmath}
\usepackage{amsthm}
\usepackage{amssymb}
\usepackage{commath}
\usepackage{mathtools}
\usepackage{enumitem}
\usepackage{booktabs}
\usepackage{algpseudocode,algorithm,algorithmicx}

\theoremstyle{plain}
\newtheorem{thm}{Theorem}

\newtheorem{defn}{Definition}

\theoremstyle{definition}

\title{Multi-Winner Contests for Strategic Diffusion in Social Networks}
\author{
Wen Shen,
Yang Feng, 
Cristina V. Lopes\\
University of California, Irvine,  California 92697, United States\\
wen.shen@uci.edu, yang.feng@uci.edu, lopes@ics.uci.edu
}

\begin{document}

\maketitle

\begin{abstract}
Strategic diffusion encourages participants to take active roles in promoting stakeholders' agendas by rewarding successful referrals. As social media continues to transform the way people communicate, strategic diffusion has become a powerful tool for stakeholders to influence people's decisions or behaviors for desired objectives. Existing reward mechanisms for strategic diffusion are usually either vulnerable to false-name attacks or not individually rational for participants that have made successful referrals. Here,  we introduce a  novel {\em multi-winner contests} (MWC) mechanism for strategic diffusion in social networks. The MWC mechanism satisfies several desirable properties, including false-name-proofness, individual rationality, budget constraint, monotonicity, and subgraph constraint.
Numerical experiments on four real-world social network datasets demonstrate that stakeholders can significantly boost participants' aggregated efforts with proper design of competitions. Our work sheds light on how to design manipulation-resistant mechanisms with appropriate contests. 
\end{abstract}

\section{Introduction}
Strategic diffusion is the process of spreading information among social media users to promote desired private or social outcomes~\cite{galeotti2009influencing}. As the impact of social media on people's daily lives continues to grow,  strategic diffusion has become a prominent tool for stakeholders (e.g., individuals, companies, governments, and NGOs) to influence people's preferences, decisions or behaviors~\cite{jackson2011diffusion,chaffey2016global}. The vast popularity of strategic diffusion in social networks is primarily because it encourages participants to take active roles in promoting stakeholders' agendas in a word-of-mouth fashion (i.e., in the forms of referrals).  This viral marketing strategy can reach a broader audience at a faster pace and are usually more economically efficient than traditional advertising such as newspapers, radios or televisions~\cite{leskovec2007dynamics,galeotti2009influencing}. 

 To unleash the power of crowds, stakeholders usually reward both direct and indirect referrals to encourage potential participants to perform the tasks early and invite influential players to participate. 
For example, if Alice refers Bob and Bob then refers Cathy, both Alice and Bob are rewarded for Cathy's purchases. However, Bob is typically given more rewards than Alice for his direct referral.  Such referral mechanisms are often called the {\em geometric reward mechanisms}, or the {\em incentive tree mechanisms}~\cite{pickard2011time}. Geometric reward mechanisms are usually effective and easy to implement~\cite{tang2011reflecting}.  As a result, they have witnessed a growing range of serious applications, including product promotion~\cite{drucker2012simpler},  disaster rescue~\cite{rahwan2013global},  global manhunt~\cite{rahwan2013global}, participatory sensing~\cite{gao2015survey} and crowdfunding~\cite{naroditskiy2014referral}. 

Despite the promising prospects of the geometric reward mechanisms, they are usually vulnerable to manipulations~\cite{drucker2012simpler}. Under many geometric reward mechanisms, indirect referrals are rewarded~\cite{pickard2011time}. A strategic player may create multiple fake accounts or identities on his behalf with one referring another  to increase his rewards. Players' such malicious behaviors are often called {\em false-name attacks}. A  false-name attack refers to a strategy that a profit-maximizing agent utilizes to gain benefits by creating multiple false identities (i.e., replicas).  False identities can be analogous to ``free riders"  in the sense that they pay less than what they should have paid. This free rider problem arises when genuine identities are unfeasible or difficult to be recognized. Unfortunately, many social network services lack effective methods to fully eliminate false identities~\cite{ferrara2016rise}. 

False-name attacks severely impede stakeholders from implementing desired individual or societal outcomes. On the one hand,  false-name attacks are undesirable because they not only diminish stakeholders' revenues but also reduce other truthful players' payoffs~\cite{drucker2012simpler}. On the other hand,  prior research indicates that  false-name attacks are pervasive in social networks since players may create false accounts with no or minimal efforts if no interventions are given~\cite{lorenz2011social,naroditskiy2014crowdsourcing,ferrara2016rise}. 

Despite that much research has been devoted to tackling the false-name-attack problems in social networks~\cite{conitzer2010false,emek2011mechanisms,todo2011false,brill2016false}, our work is most closely related to the mechanism design problem for multi-level marketing~\cite{drucker2012simpler}. In their work,  \citeauthor{drucker2012simpler} introduce a class of mechanisms by capping the rewards a player can get from indirect referrals.  They show that under their mechanisms false-name attacks are unprofitable. While illuminating, their methods indicate that some of the less influential players in the referral networks will receive no rewards for successful indirect referrals.  For instance, players with ordinary or low capabilities may have little or no incentives to participate in the referral mechanisms.  Ignoring these players is problematic for stakeholders because the population of many online platforms typically consist of a  large portion of participants with low or common capabilities~\cite{ipeirotis2010analyzing,dow2012shepherding}.  Besides, the implementation of their mechanisms is not based on graphs but rather on trees, it remains unknown whether their mechanisms can be applied to large-scale social networks consisting of  thousands of participants with different abilities. 

To address these issues, we introduce a novel mechanism called the {\em Multi-Winner Contests} (MWC) mechanism for strategic diffusion in social networks. The MWC mechanism distinguishes itself from existing methods in two aspects. First, it allocates virtual credits for each successful referrals by taking both diffusion contributions and verifiable task efforts into account. Second, it determines the diffusion rewards for each player according to the results of contests that compare the virtual credits earned by the qualified players. We show that the mechanism is false-name-proof, individually rational, budget-constrained, monotonic, subgraph-constrained, and computationally efficient. We conducted extensive experiments with four real-world social network datasets. Experimental results demonstrate that false-name-attacks are unprofitable under the MWC mechanism. Stakeholders can significantly boost the aggregated efforts of players when they select parameters of the MWC mechanism appropriately. Our work casts light on how to integrate competitions into the design of novel mechanisms to counter manipulations.

\section{Strategic Diffusion in Social Networks}
This section first describes notations used for modeling strategic diffusion in social networks. It then introduces the mechanism design problem for strategic diffusion. After defining the concept of false-name attacks, it presents the solution concepts for the reward mechanism design problem.
\subsection{Preliminaries}
\begin{figure}
\centering
 \includegraphics[width=.30\columnwidth]{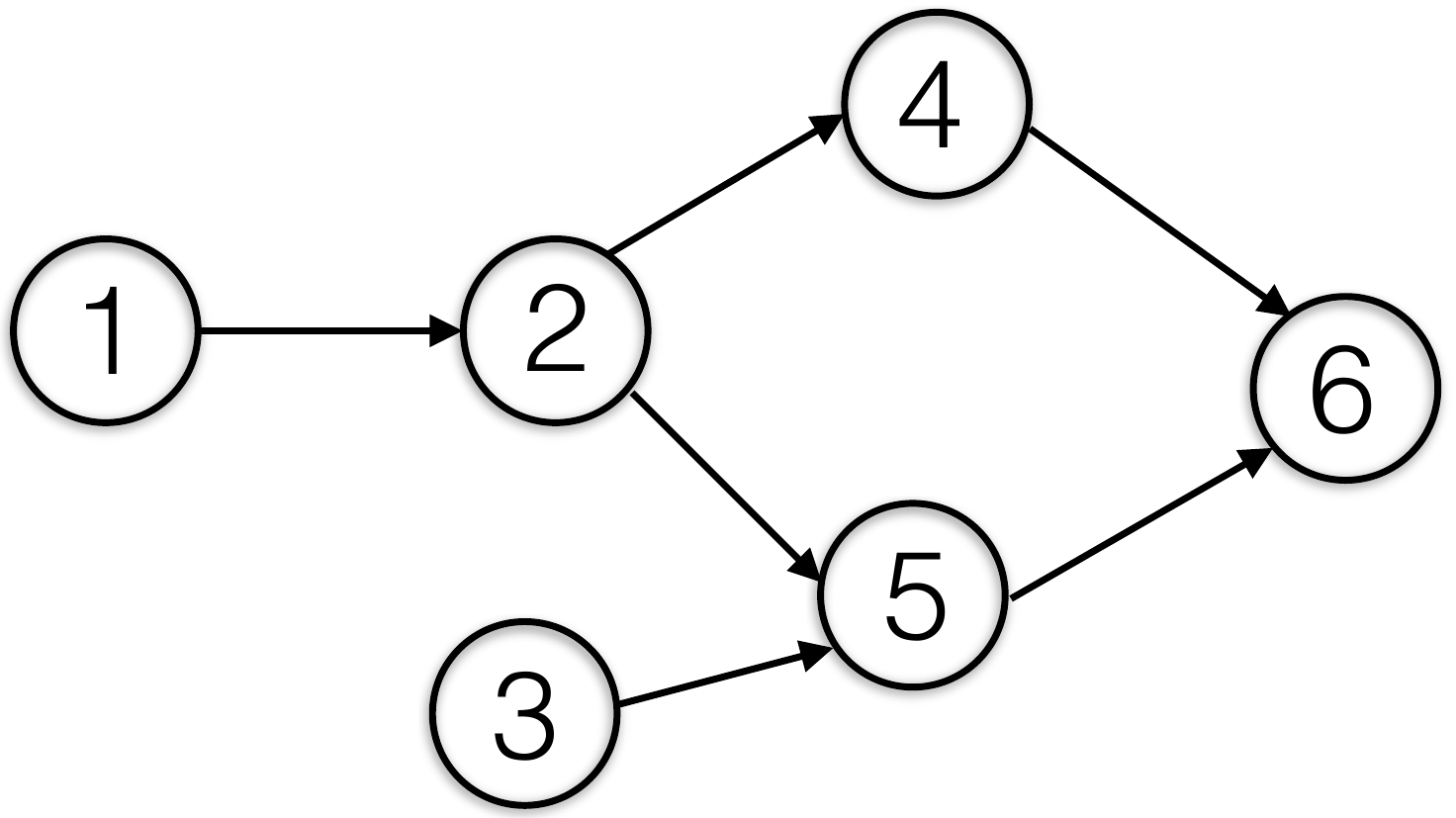}
\caption{A typical directed acyclic graph.}
\label{fig:diffdemo}
\end{figure}

Strategic diffusion processes or referral networks are usually modeled with {\em directed acyclic graphs} (DAGs).  We consider a referral DAG $G = (V, E)$ (See Figure~\ref{fig:diffdemo}) where $V$ denotes the set of players that may contribute (i.e., nodes) and $E$ denotes the set of referral relationships (i.e., edges).  For any nodes $v, u \in G$ ($v\neq u$), if  there is a directed edge from $v$ to $u$, then it means that $u$'s decision to contribute is partially a result of $v$. In this case, we say $u$ is a {\em direct successor} of $v$ and $v$ is a {\em direct predecessor} of $u$. For each node $v$, the number of direct predecessors $v$ has is its {\em indegree} $deg^{-}(v)$. The number of direct successors it has is its {\em outdegree}, denoted by $deg^{+}(v)$.  A source node (i.e., a seed node) has a indegree of 0, while a sink node has a outdegree of 0. 

For any nodes $v, u \in G$ ($v\neq u$), if a path leads from $v$ to $u$, then $u$ is said to be a  {\em successor} of $v$ and reachable from $u$, and $v$ is said to be a {\em predecessor} of $v$. We write the distance  between $v$ and $u$ by $dist(u,v)$. If there is no path from $v$ to $u$, then the distance between the two nodes is infinity; that is $dist(v,u) = \infty$. The length from $v$ to $v$ is $0$, i.e., $dist(v,v) = 0$.  Let $\kappa^{+}_{v}$ be the set of all successors of $v$, we have $\kappa^{+}_{v} = \{ u \in V :  0<dist(v, u)< \infty \}$.  Similarly, the set of $v$'s predecessors $\kappa^{-}_{v} = \{ u \in V :  0 < dist(u,v) < \infty\}$.  Let $G_{v}$ be the subgraph rooted at $v$, then $G_{v} =  (V', E')$, where $V' = \{v\} \cup \kappa^{+}_{v}$ and $E' \subseteq E$.  

In a DAG $G = (V, E)$, if  a player $v \in V$ makes a contribution to the designated tasks (e.g., making purchases, answering questions, reporting software bugs), we say player $v$ exerts {\em task efforts} $t_{v} \in \mathbb R_{\ge 0}$.  Player $v$ may either directly or indirectly  spread the information  of the tasks to his successors  to maximize his profits.  If one of his successors $u$ also contributes to the tasks, then we say $v$ makes {\em diffusion contributions} $d_{v} \in \mathbb R_{\ge 0}$. In our model, we assume that the task efforts are verifiable and available for the stakeholder. For instance, a seller usually has the legitimate transactions information about the products she sells to each buyer. 

Each edge $e = (v_{i}, v_{j}) \in E$ has a weight $\omega_{v_{i}v_{j}}$ (i.e., $\omega_{e}$). The weight $\omega_{v_{i}v_{j}}$ represents the proportion of the credits assigned to $v_{i}$ when $v_{j}$ contributes to the designated tasks with efforts $t_{v_{j}}$. Here, $\omega_{v_{i}v_{j}} =  1/deg^{-}_{v_{j}}$ if $v_{i} \neq v_{j}$; otherwise, $\omega_{v_{i}v_{j}}  = 1$. Let $P_{v_{i}v_{j}}$ be the set of all paths from $v_{i}$ to $v_{j}$ and $p$ be a path in the set $P_{v_{i}v_{j}}$. Path $p$'s weight $\omega(p)$ is the product of the weights of all edges along the path.  That is,  $\omega(p) = \prod_{(v_{i}, v_{j})\in p} \omega_{v_{i}v_{j}}$. Here, the length of the path $|p| = dist(v_{i}, v_{j})$.

\subsection{Mechanism Design for Strategic Diffusion}
We consider a principal (e.g., a seller, a task owner)  employs strategic diffusion to maximize the aggregated efforts on the tasks she designates. Initially, the principal selects some participants (i.e., seeds) $S \subseteq V$ in a social network $G = (V, E)$ to perform the tasks that she specifies. The principal may select the seed nodes randomly if she has no prior knowledge of the social network. Alternatively, she may select the seed nodes using the influence maximization approach if  the network structure and diffusion method are known a priori~\cite{kempe2003maximizing} . 

Some of the seed participants then perform the tasks and invite their neighbors in the social network to participate. If a participant $v$ exerts task efforts $t_{v}$, he will receive a task reward $\pi_{t}(v)$ from the principal. Participant $v$ may spread the information to his successors $u \in \kappa^{+}_{v}$ to maximize his profits. If a successor $u$ exerts task efforts $t_{u}$,  the referrer $v$ will receive a diffusion reward $\pi_{d}( v, u)$. The total diffusion rewards for $v$: $\pi_{d}(v) = \sum_{u \in \kappa^{+}_{v}} \pi_{d}(v, u)$. Thus, the total rewards for $v$ are determined by: $\pi (v) = \pi_{t}(v) + \pi_{d}(v)$.

The total contributions of $v$ include both the task efforts and the diffusion contributions. In strategic diffusion, the task efforts are verifiable while the diffusion contributions are difficult to verify because strategic players can generate fake referrals with no or minimal efforts by creating multiple false identities with one being referred by another. We thus treat them separately.  Let $c(v)$ be the total contributions for $v$, we have: $c(v) = (t_{v},  d_{v}) $.  
Let $\Theta_{v}$ be the  space of $v$'s total contributions $c(v)$, and $\Theta = (\Theta_{v})_{v\in V}$. 

In our model, we assume that the valuation of a player is linear in the rewards  he receives and the cost he pays is linear in the task efforts he has contributed. That is, there are constants $\alpha_{v}$, $\beta_{v}$, $\gamma_{v}$ and $\varsigma_{v}$ with $\beta_{v} > 0$ and $\varsigma_{v}> 0$, such that player $v$'s utility is $U (v) := \alpha_{v} + \beta_{v} \cdot \pi(v) - (\gamma_{v} + \varsigma_{v} \cdot t_{v})$. Without loss of generality,  let $\alpha_{v} = \gamma_{v}$ and $\delta_{v} = \frac{\varsigma_{v}}{\beta_{v}}$. We have player $v$'s  utility:
\begin{equation}
\label{eq:utility}
U(v) = \pi(v)- \delta_{v} \cdot t_{v} \;,
\end{equation}
 where $\delta_{v} > 0$ is a private coefficient that determines the player's marginal cost for exerting extra unit effort. The higher ability a player owns, the lower $\delta_{v}$ he has, and vice versa.  That is, there is a negative correlation between  a player's ability and his cost coefficient. 
 
 We note that  a player may incur costs on spreading the information to his neighbors. We avoid explicitly including players' diffusion costs into the utility function for two concerns. First, diffusion efforts are not directly verifiable for the principal. The exact correlations between the costs on task efforts and the diffusion costs may vary from player to player, which are unknown to the principal. Second, a player $v$ can integrate his diffusion costs into the task efforts by setting a higher cost coefficient $\delta_{v}$. Similar techniques were also used in literature~\cite{shen2018information}.
 
In strategic diffusion, the principal is interested in a {\em reward mechanism} $\pi$ that determines the reward for each player that has exerted efforts. 
\begin{defn}
A reward mechanism $\pi$ is a tuple of payments for each player $v \in V$, where $G= (V,E)$.  That is, $\pi = (\pi(v))_{v \in V}$, where $\pi(v): \Theta \rightarrow \mathbb{R}$.
\end{defn}

\subsection{False-Name Attacks}
In a graph $G = (V, E)$, we say a graph $G' = (V'', E'')$ and a set of replicas $R \subseteq V'$ are a {\em false-name attack} by $v$ in $G$ if  when we collapse $R$ into the single node with label $v$ in $G'$ we get the graph $G$. 
\begin{defn}[False-Name Attack]
\label{defn:falsenameattack}
Given a referral graph $G = (V, E)$,  for any $v \in V$, let $DS_{v}$, $DP_{v}$ be the sets of node $v$'s direct successors and direct predecessors, respectively. $G' = (V'', E'')$ is obtained from $G$ by a false-name attack at $v$ if:
\begin{enumerate}
\item[ $\;\bullet$ ] $V'' = V \setminus \{ v\} \cup \{ r_{1}, ..., r_{m}\}$, $m > 1$. The set of nodes $R = \{r_{1}, ..., r_{m}\}$ is the set of replicas of $v$.
\item[ $\;\bullet$ ] The sum of the task efforts of $v$'s replicas is equal to $v$'s task efforts. That is,  $t_{v} = \sum_{r \in R} t_{r}$, where $t_{r} > 0$.
\item[ $\;\bullet$ ] All replicas of $v$ have at least one direct successor; that is $deg^{+}(r) \geq 1$ for all $r \in R$.
\item[ $\;\bullet$ ] For all $u \in DS_{v}$, $G_{u} = G'_{u}$, where $G'_{u}$ is a subgraph of $G'$ rooted at $u$.
\item[ $\;\bullet$ ] The direct predecessors of a replica of $v$ are either replicas of $v$, or the direct predecessors of $v \in V$, or null. That is,  $DP_{r} = R \setminus \{r \} \cup DP_{v} \cup \emptyset$, where $r \in R$.
\item[ $\;\bullet$ ] $\forall r \in R, \exists $ a replica $r'$ of $v$ such that $r \in \kappa^{+}_{r'}$.  
\end{enumerate}
\end{defn}

\begin{figure}
\centering
 \begin{subfigure}[b]{.48\columnwidth}
  %\centering
  \includegraphics[width=\columnwidth]{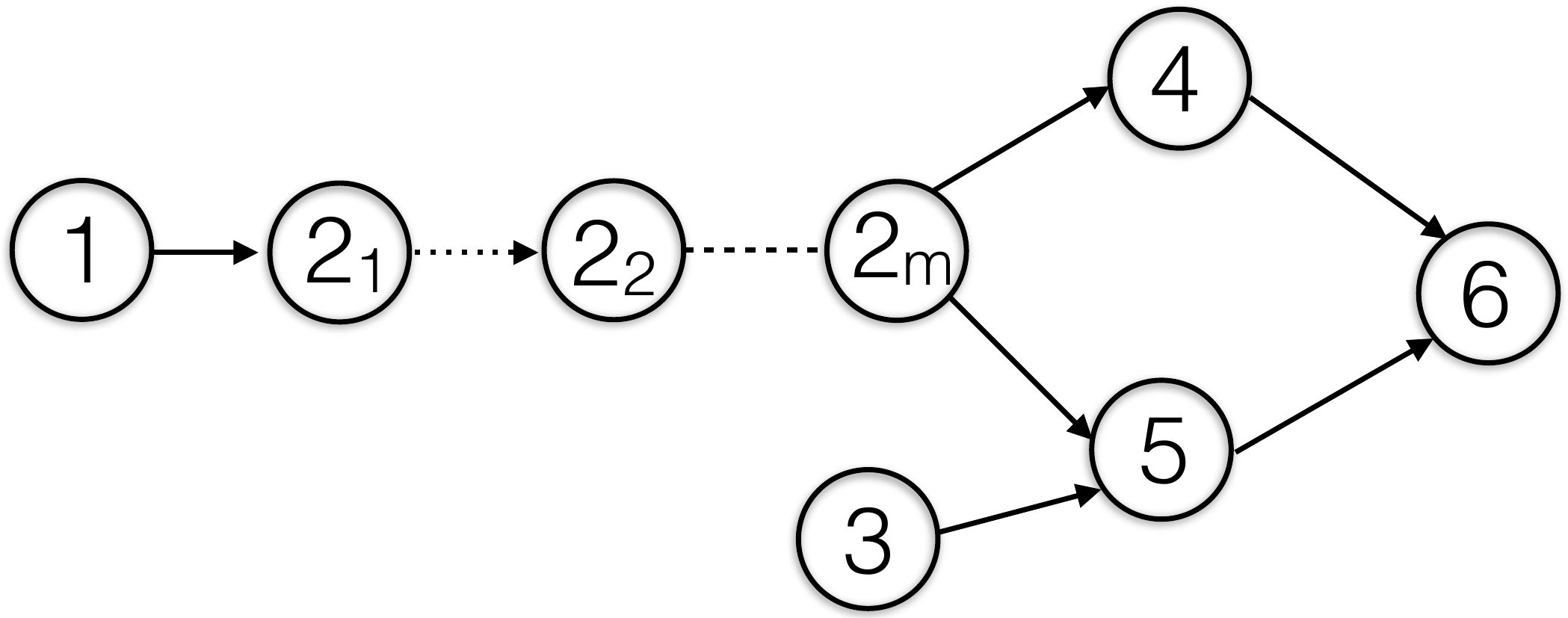}
  \caption{Type 1}
  \label{fig:sybil1}
\end{subfigure}
\hfill
\begin{subfigure}[b]{.42\columnwidth}
  %\centering
  \includegraphics[width=\columnwidth]{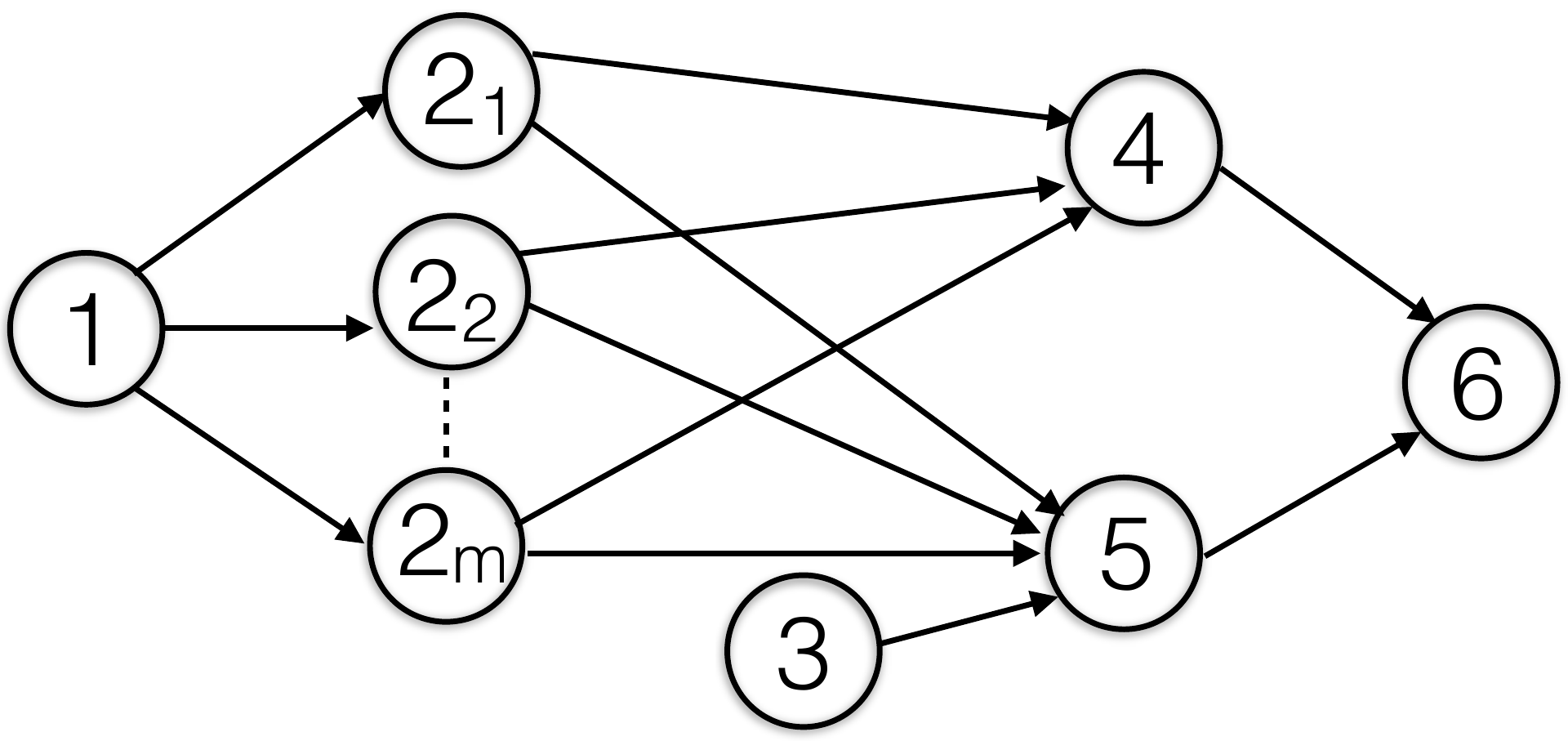}
    \caption{Type  2}
      \label{fig:sybil2}
\end{subfigure}

\caption{False-name attacks at node 2: nodes $2_{1}, 2_{2}, ..., 2_{m}$ are replicas of node 2.}
\label{fig:sybildemo}
\end{figure}

In general, there are three types of false-name attacks: type 1 (See Figure~\ref{fig:sybil1}), type 2 (See Figure~\ref{fig:sybil2}), and hybrid. Type 1 false-name attacks occur in the form of a long referral chain that consists of the replicas, while in type 2 attacks the replicas operate in parallel. Hybrid attacks are combinations of type 1 and type 2 attacks. 

A false-name attack is profitable if the sum of the rewards received by the replicas $r \in R$ of $v \in V$ are higher than the reward  $v$ receives. That is, $\sum_{r \in R} \pi(r) >\pi(v) $.
\subsection{Solution Concepts}

Since false-name attacks are harmful to both the stakeholders and other truthful players, it is desirable that the reward mechanism satisfies the  {\em false-name-proofness} property. A reward mechanism is false-name proof if false-name attacks are unprofitable for every player in a social network. That is, the total rewards that $v \in G$ receives are at least the same as the sum of the rewards received by the replicas of $v$ in a graph obtained from a split of $G$ at node $v$. 
\begin{defn}[False-Name-Proofness]
A reward mechanism $\pi$ is false-name-proof if for all $v \in G$: $\pi (v) \geq \sum_{r \in R} \pi (r)$, where $R$ is the set of replicas due to a false-name attack at node $v$.
\end{defn}

 According to Equations~\ref{eq:utility} and~\ref{eq:totalreward}, if player $v$'s costs on task efforts (i.e., $\delta_{v} \cdot t_{v}$) are equal to or higher than the rewards $\pi_{t} (v)$, then player $v$ will have no incentives to participate in the strategic diffusion without positive diffusion rewards. To encourage potential players to participate, a reward mechanism needs to ensure that a player who has made successful referrals receives positive (expected) rewards. Otherwise, it would be not appealing to players with high marginal costs (i.e., high $\delta_{v}$).  A reward mechanism satisfies  {\em individual rationality} if its (expected) utility is positive and it allocates positive (expected) diffusion rewards to each player that have made successful referrals.
\begin{defn}[Individual Rationality]
A reward mechanism $\pi$ is individually rational if for each player $v \in V$ with $d_{v} > 0$, we have: $U(v) >0$, and $\pi_{d}(v, u) > 0$,  where $u \in \kappa^{+}_{v}$, and $U(v)$ is determined by Equation~\ref{eq:utility}.
\end{defn}

In practice, a reward mechanism should be budget constrained. If not, the mechanism will be economically unfeasible for deployment. In our setting, the total rewards distributed to the players should not exceed a fixed portion $\vartheta$ of the total aggregated task efforts. 
\begin{defn}[Budget Constraint]
A reward mechanism $\pi$ is budget constrained if: $\sum_{v \in G} \pi(v) \leq \vartheta \cdot \sum_{v \in G} t_{v}$, where $\vartheta$ is a positive constant.
\end{defn}

Another important constraint is that a reward mechanism should limit indirect referrals to restrict the formation of long referral chains. This property is usually called the {\em monotonicity} constraint. A reward mechanism is monotonic if direct referrals receive higher diffusion rewards than or at least the same rewards as indirect referrals. The monotonicity property encourages participants to form short diffusion chains by offering more rewards. As a result, it limits the scope of indirect rewards, which is desirable in practice because many successful  referral chains are usually short~\cite{leskovec2007dynamics}.
\begin{defn}[Monotonicity]
A reward mechanism $\pi$ is monotonic If $v_{2}$ is a successor of $v_{1}$, adding a direct successor $v_{i}$ to $v_{2}$ increases $v_{1}$'s diffusion rewards $\pi_{d}(v_{1}^{'})$ at least as much as the diffusion rewards  $\pi_{d}(v_{1}^{''})$ by adding a direct successor $v_{j}$ to a successor of $v_{2}$ ,  where $t_{v_{i}} = t_{v_{j}}$, i.e., $\pi_{d}(v_{1}^{'}) \geq \pi_{d}(v_{1}^{''})$.
\end{defn}

In strategic diffusion, players should not have incentives to delay performing the tasks to wait for a referral with a more rewarding position in a social network $G$. To satisfy this constraint, a reward mechanism should determine the rewards for each player $v \in V$ based on the subgraph $G_{v}$ rooted at $v$.
\begin{defn}[Subgraph Constraint]
A reward mechanism $\pi$ is subgraph-constrained if $\pi (v)$ only depends on the rooted subgraph $G_{v}$.
\end{defn}
\section{Multi-Winner Contests Mechanism}
We present  a novel reward mechanism called the {\em Multi-Winner Contests} (MWC) mechanism. It has two key ingredients. The mechanism first calculates the virtual credits for the diffusion contributions of each player with successful referrals. It then determines the diffusion rewards by holding a contest among players that are in his rooted subgraph. The MWC mechanism is computationally efficient and satisfies several desirable properties, including false-name-proofness, individual rationality, budget constraint, monotonicity, and subgraph constraint.

\subsection{Virtual Credits for Diffusion Contributions}
For each newly joined player $v$ that has exerted task efforts $t_{v}$, the MWC mechanism pays $\pi_{t}(v) = \mu \cdot t_{v}$ ($\mu >0$) for his task contributions and allocates virtual credits $\eta \cdot (t_{v})^{2}$ for his diffusion contributions. If $v$ has  either directly or indirectly referred  player $u \in \kappa_{v}^{+}$ to participate ($t_{u} > 0$), his virtual credits $b_{v}$ are computed by:
\begin{equation}
\label{eq:virtualcredits}
b_{v}= \eta \cdot (t_{v})^2 + t_{v} \cdot  \sum_{u \in \kappa^{+}_{v}} \sum_{p \in P_{vu}}  t_{u} \cdot \omega(p) \cdot  \lambda^{|p|} \; ,
\end{equation} 
where $0<\lambda < 1$,  $\eta \geq \lambda/2$, and $P_{vu}$ is the set of  paths from $v$ to $u$. 

The initial allowance of virtual credits  $\eta \cdot (t_{v})^{2}$  serves  dual purposes. First, it ensures that each player has a positive amount of virtual credits to enter into the contests and receives a positive share of diffusion rewards. Second, it allows the mechanism to significantly reduce a player's virtual credits if he splits his task efforts. This dedicated design is an essential step that makes false-name attacks unprofitable.  Note that $t_{u} \cdot \lambda^{|p|}$ is a typical {\em tree incentive mechanism} (i.e., geometric reward mechanism). The MWC mechanism extends it to a reward mechanism that applies to social networks by introducing the weights $w(p)$. To incentivize players to exert higher task efforts, the MWC mechanism amplifies the virtual credits for diffusion contributions by multiplying the player's task efforts $t_{v}$. 

\subsection{Diffusion Rewards}
The MWC mechanism performs multi-winner contests to decide the diffusion rewards. In the contests, the winners are chosen simultaneously by a {\em contest success function} (CSF) that compares the virtual credits of each player.  A CSF determines each player's probability of winning the contest in terms of all players' efforts~\cite{skaperdas1996contest}.

In general, there are two types of contest success functions: the {\em ratio} form and the {\em difference} form~\cite{skaperdas1996contest}. In the ratio form, the winning probabilities depend on the ratio of efforts exerted by each player.  In the difference form,  they  are determined by the difference in efforts that each player has exerted. The ratio form can be naturally applied to large contests that consist of many players, while extending the difference form to contests with more than two players is usually non-trivial and difficult~\cite{jia2013contest}. Strategic diffusion in social networks typically involves a large number of players. It is thus desirable to use the ratio form  CSFs for analytical convenience. 

The ratio form CSFs typically predict contest outcomes from the ratio of the efforts that each player has devoted. For each node $v \in G$, the MWC mechanism holds a contest for each subgraph $G_{v} = (V', E')$ rooted at $v$.  If player $v$'s virtual credits $b_{v}  \geq \eta \cdot (t_{v})^2$, he will be allowed to enter into the contests. Otherwise, player $v$ makes no  contributions and receives 0  rewards. If $v$ is allowed to enter into the contest, his probability of winning is determined by a contest success function: $prob(v) = \frac{(b_{v})^{\sigma}}{\sum_{u \in V'} (b_{u})^{\sigma}}$, 
where $0<\sigma \leq 1$,  and $b_{v}\geq \eta \cdot (t_{v})^2$. 

The parameter $\sigma$ can be interpreted as the ``noise" of a contest. It captures the marginal increase in the probability of winning caused by a higher effort and is crucial to the outcomes of the contest~\cite{jia2013contest}. Contests with low $\sigma$ can be regarded as poorly discriminating or  ``noisy" contests. That is, players with different efforts may have a similar level of chance to win.  Contests with high $\sigma$ can be regarded as highly discriminating in the sense that players with higher efforts have a greater chance to win. The contest success functions are imperfectly discriminating in the sense that with all of them, the prize at stake is awarded probabilistically to one of the players with higher virtual credits leading to a higher probability of winning the prize. 

In the contest success function, each player's probability of success does not depend on his identity or the identities of his opponents, but just on the efforts (i.e., virtual credits) of the players. This property indicates that if in some cases two players have identical efforts then their probabilities of success must be equal and if all players were to exert identical efforts, then each one of them would have a probability of success equal to $1/ n$. Besides, a player's probability of success is independent of agents who have not exerted efforts.

The total rewards for diffusion contributions are $\pi_{d} = \phi \cdot  \sum_{u \in V'} t_{u} \frac{deg^{-}_{G_{v}} (u)}{deg^{-}(u)}$, where  $\phi > 0$, and $deg^{-}_{G_{v}} (u)$ denotes the number of direct predecessors of $u$ in $G_{v}=(V', E')$. Here, $deg^{-}_{G_{v}} (u) \leq deg^{-}(u) $.
The diffusion rewards of $v$ are thus determined as:
\begin{equation}
\label{eq:diffreward}
\pi_{d}(v) = \frac{(b_{v})^{\sigma}}{\sum_{u \in V'} (b_{u})^{\sigma}} \cdot \phi \cdot  \sum_{u \in V'} t_{u} \cdot \frac{deg^{-}_{G_{v}} (u)}{deg^{-}(u)}\;.
\end{equation}
The MWC mechanism utilizes the probability of winning to determine the proportion of the total diffusion rewards for each player.  This is because empirical evidence shows that proportional-prize contests exert higher aggregated efforts than winner-take-all contests~\cite{cason2010entry,sheremeta2011contest}. Besides, proportional-prize contests also limit the degree of biases that discourage low-ability players without altering the performance of stronger players~\cite{cason2010entry,Cason2018winner}.
\subsection{The MWC Mechanism}
 The MWC mechanism $\pi = (\pi(v))_{v\in G}$ uses a post-price mechanism $\pi_{t}(v) = \mu \cdot t_{v}$ to reward a player that exerts $t_{v}$ task efforts. The total reward for player $v$ includes both the task rewards $\pi_{t}(v)$ and the diffusion rewards $\pi_{d}(v)$. Thus, we have the total reward for player $v$:
\begin{equation}
\label{eq:totalreward}
\pi (v)= 
\begin{cases}
\mu \cdot t_{v} + \pi_{d}(v) & \quad \text{if } b_{v} \geq \eta \cdot (t_{v})^{2}\\
0 & \quad \text{otherwise} \;,
\end{cases}
\end{equation}
where $\mu \in \mathbb R_{> 0}$ is the reward parameter that characterizes to what extent the principal values agents' efforts, and $b_{v}$, $\pi_{d}(v)$ are defined by Equations~\ref{eq:virtualcredits} and~\ref{eq:diffreward}, respectively.

The MWC mechanism $\pi$ (See Algorithm~\ref{alg:mwc} for an implementation)  is computationally efficient. The social network $G = (V, E)$ can be implemented using adjacency lists, where each node maintains a list of all its adjacent edges. It takes $O(|V|^2+|V|\cdot |E|))$  time to compute the virtual credits (See Lines 1--6).  Computing the total rewards (See Lines 7--9) also takes  $O(|V|^2+|V|\cdot |E|))$  time. Therefore, the time complexity for the MWC mechanism is $O(|V|^2+|V|\cdot |E|))$.

\begin{algorithm}[ht]
  \caption{{\em Multi-Winner Contests Mechanism}
    \label{alg:mwc}}
  \begin{algorithmic}[1]
\Statex \textbf{Input: }$G$- a social network; $S$-seed nodes
 \Statex \textbf{Output: } $\pi $-payment for each node $v \in G $
\Statex  \textbf{Initialize:} $\pi(v) = 0$ for all $v\in G$
\For{each newly joined player $v \in G$  {\bf and} $t_{v} > 0$}
\State $b_{v} \gets \eta \cdot (t_{v})^{2}$
\For{each $v_{i} \in \kappa^{-}_{v}$ {\bf and} $t_{v_{i}} > 0$}
\State $b_{v_i} \gets b_{v_i} + t_{v_{i}} \cdot t_{v} \cdot \sum_{p \in P_{v_{i}v}} w(p) \cdot \lambda^{|p|}$
\EndFor
\EndFor
\For{each $v_{j}$ in $G_{v}$}
\State Compute the rewards $\pi(v_{j})$ by Equation~\ref{eq:totalreward}
\EndFor\\
\Return $\pi$
  \end{algorithmic}
\end{algorithm}

The MWC mechanism  achieves false-name-proofness by introducing an initial allowance $\eta_{v} \cdot (t_{v})^2$ of virtual credits for each player $v$ that has exerted non-zero task efforts. If the task efforts reduce, the allowance will decrease quadratically. False-name-attacks are not profitable because splitting task efforts into smaller pieces causes a larger degree of reduction than the degree of increase in virtual credits. This elegant design of the virtual credits enables the MWC mechanism to be manipulation-resistant to all the three types of false-name attacks. We postpone all the proofs to the Appendix.
\begin{thm}[False-Name-Proofness]
\label{thm:fnp}
The MWC mechanism $\pi$  is false-name-proof.
\end{thm}

The MWC mechanism employs a post-price mechanism $\pi_{t}$ to determine the rewards for players' task efforts. For the diffusion rewards, each player that has made successful referrals is guaranteed to receive a positive amount of rewards. Thus, it follows that the MWC mechanism satisfies the individual rationality property.
 
\begin{thm}[Individual Rationality]
The MWC mechanism $\pi$ is individually rational. 
\end{thm}

The MWC mechanism is budget-constrained because it allocates a fixed portion of rewards (i.e., $\mu \cdot \sum_{v \in G} t_{v}$) for the task efforts and a fixed number of rewards (i.e., $\phi \cdot \sum_{v \in G}t_{v}$) for diffusion rewards.  
\begin{thm}[Budget Constraint]
The MWC mechanism $\pi$ is budget-constrained.
\end{thm}

Under the MWC mechanism, players with direct referrals are given more virtual credits than if they have indirect referrals.  This is achieved by introducing the discounting factor $\lambda < 1$. For the same level of task efforts, the virtual credits for diffusion rewards decrease as the distance of path between the referrer and the successor increases.
\begin{thm}[Monotonicity]
The MWC mechanism $\pi$ is monotonic.
\end{thm}

Since the diffusion rewards are determined by a contest among players in a subgraph rooted at player $v$, it follows that the MWC mechanism is subgraph constrained.
\begin{thm}[Subgraph Constraint]
The MWC mechanism $\pi$ satisfies the subgraph constraint.
\end{thm}

\section{Experiments} 
Before describing the results, we introduce the experimental settings.
\subsection{Experimental Settings}
We used four publicly available datasets: Twitter~\cite{hodas2014simple}, Flickr~\cite{cha2009measurement}, Flixster~\cite{goyal2011data},  and Digg~\cite{hogg2012social}. These datasets included anonymized timestamps that could be used to estimate the influence probabilities needed  for simulating the diffusion process of social networks~\cite{goyal2010learning}. For each dataset, we first estimated the influence diffusion probabilities for each node using the learning algorithms by~\citeauthor{goyal2010learning} (\citeyear{goyal2010learning}) with the Bernoulli distribution under the static model. We then simulated the influence diffusion process with the general threshold model~\cite{kempe2003maximizing} based on the estimated diffusion probabilities. After preprocessing, each dataset produced a largest weakly connected component.  See Table~\ref{tab:dataset} for dataset configuration. 
\begin{table}[h]
\centering
\resizebox{0.9\columnwidth}{!}{
\begin{tabular}{lrrrrr} \toprule
    {Dataset} & {$\#$Nodes} & {$\#$Edges} & {$\#$Seeds} & {M.D.} & {A.D.}  \\ \midrule
    Twitter  & 323,185 & 2,148,717 & 1,715 & 8,822 & 52 \\ \midrule
     Flickr  & 145,305  & 2,149,882 & 768  & 6,731  & 34   \\ \midrule
    Flixster  & 95,969  & 484,865 & 502  & 3,109 & 27 \\ \midrule
    Digg & 17,817  & 128,587 & 107  & 1,375  & 20\\ \bottomrule
\end{tabular}
}
\caption{Dataset configuration: M.D. -- maximum degree, A.D. -- average degree.}
\label{tab:dataset}
\end{table}

\begin{figure}
\centering
\begin{subfigure}[b]{.38\columnwidth}
  %\centering
  \includegraphics[width=\columnwidth]{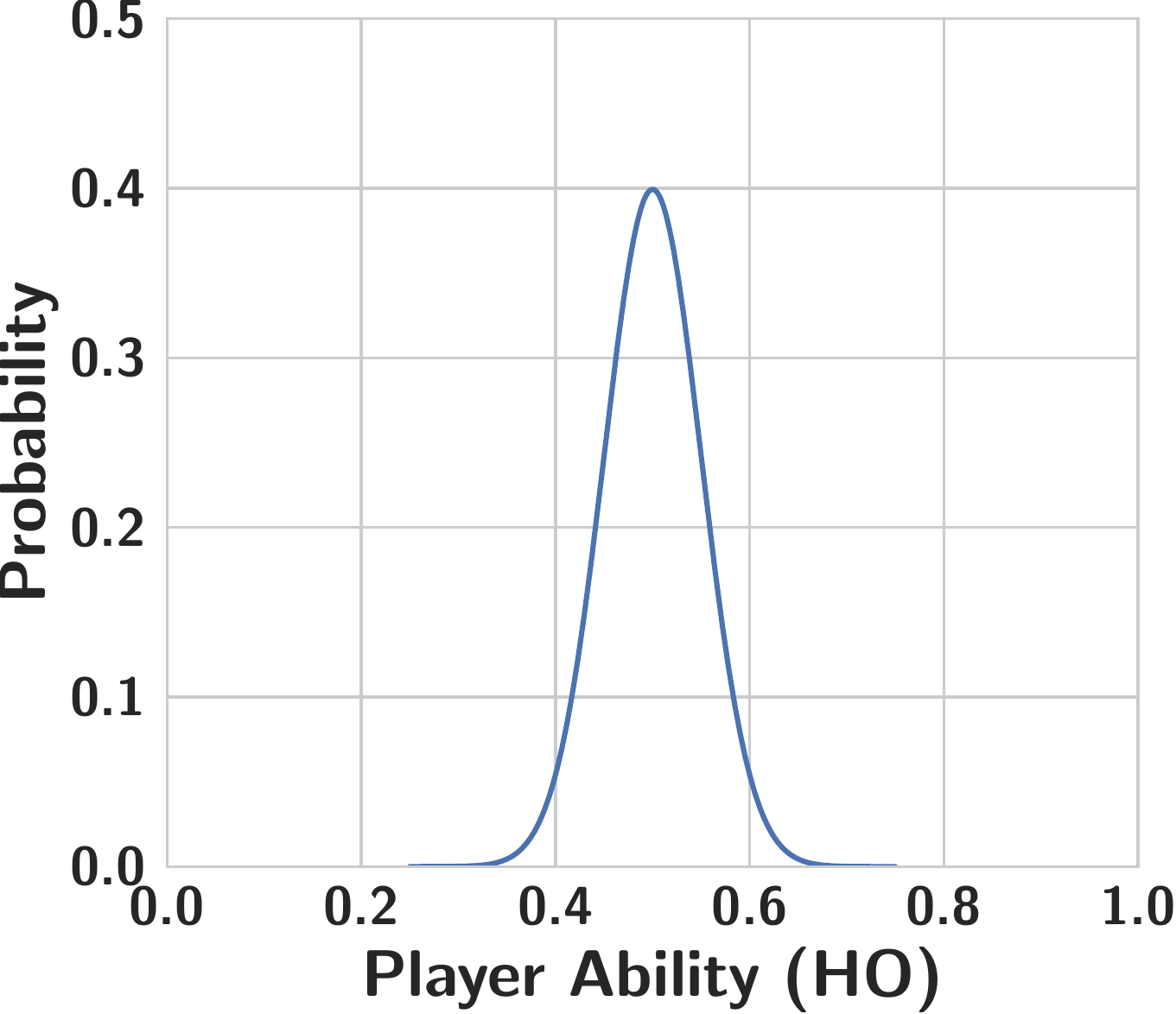}
  \caption{mean=.5, std=.05}
  \label{fig:hopdf}
\end{subfigure}
\begin{subfigure}[b]{.38\columnwidth}
  %\centering
  \includegraphics[width=\columnwidth]{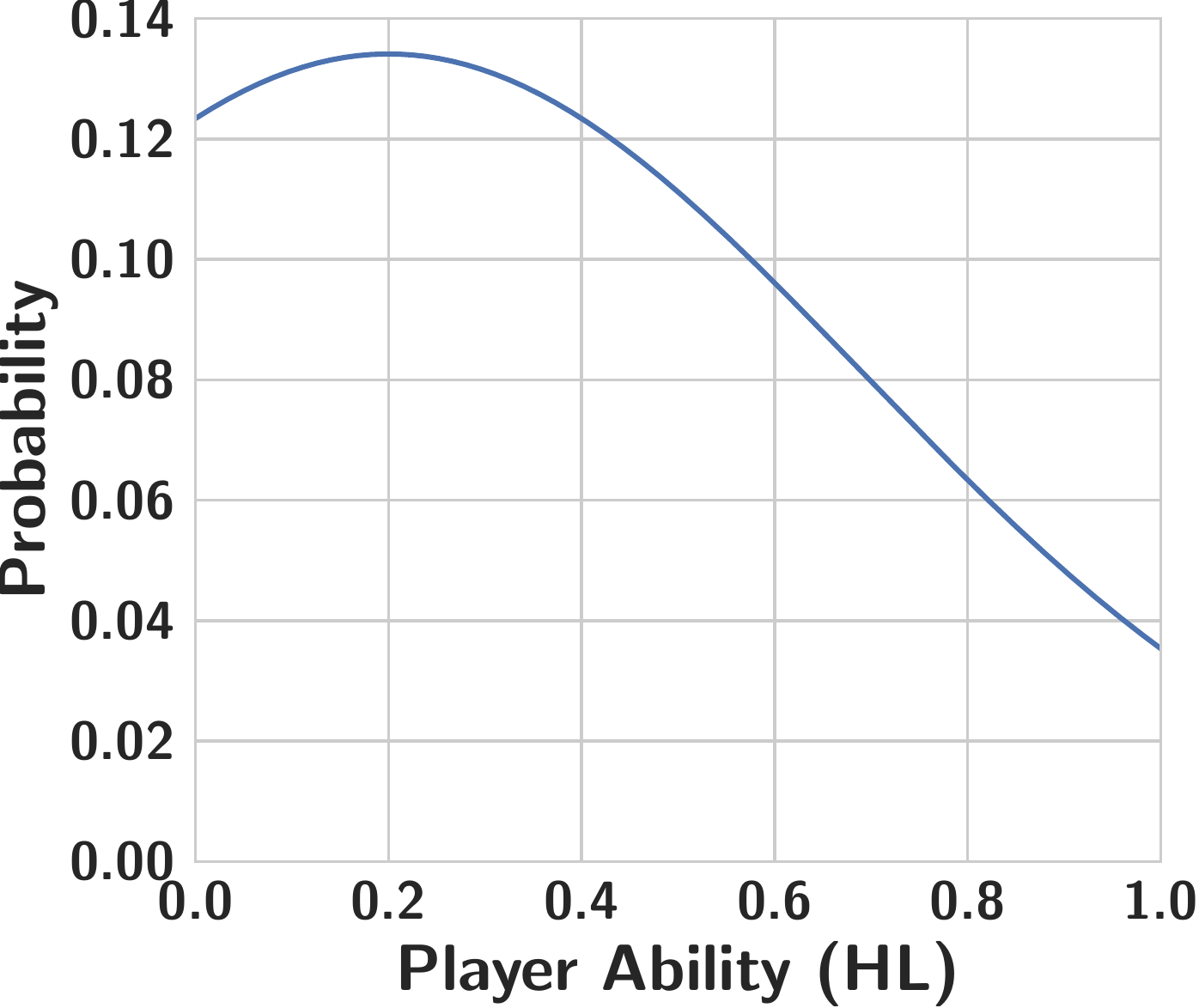}
    \caption{mean=.2, std=.7}
      \label{fig:hlpdf}
\end{subfigure}
\begin{subfigure}[b]{.38\columnwidth}
  %\centering
  \includegraphics[width=\columnwidth]{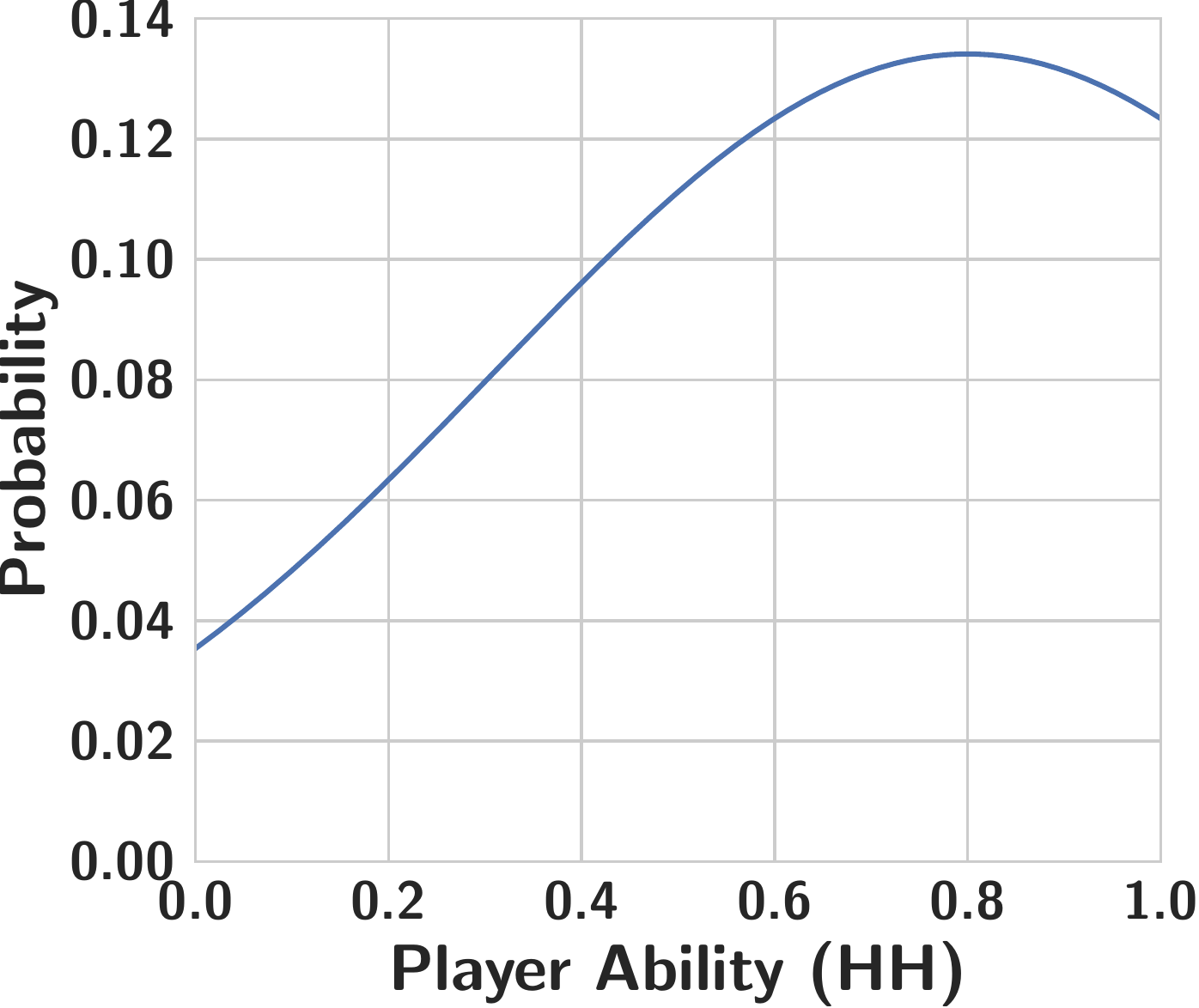}
    \caption{mean=.8, std=.7}
     \label{fig:hhpdf}
\end{subfigure}
\begin{subfigure}[b]{.38\columnwidth}
  %\centering
  \includegraphics[width=\columnwidth]{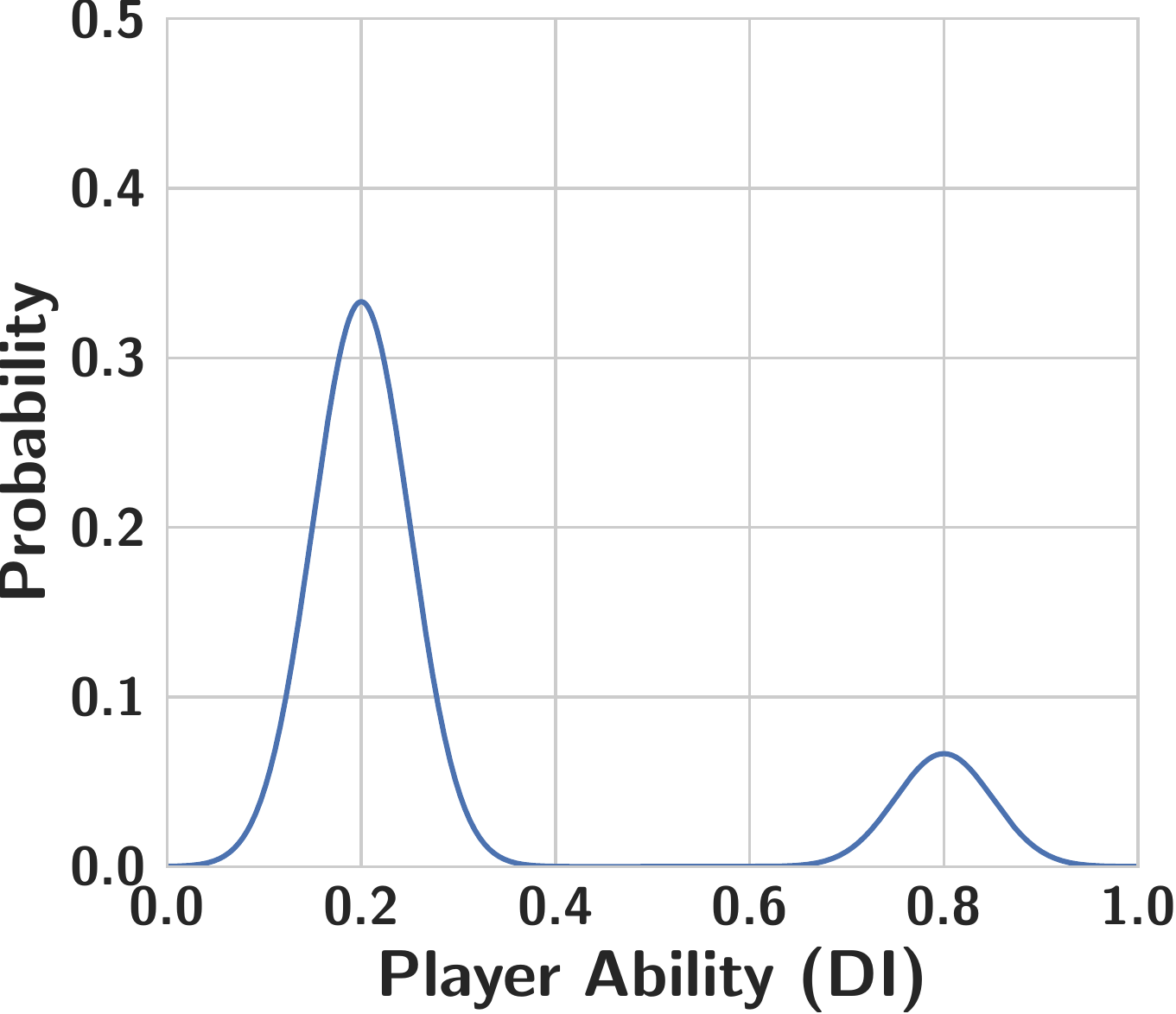}
    \caption{means=.2, .8, std=.05}
     \label{fig:dipdf}
\end{subfigure}
\caption{Probability density function of the  means of players' abilities by four different groups: HO, HL, HH and DI.}
\label{fig:pdfs}
\end{figure}

We modeled players' abilities  with the simulation method by~\citeauthor{burnap2013simulation} (\citeyear{burnap2013simulation}). We considered four groups of players: {\em homogeneous (HO)}-players with similar levels of abilities; {\em heterogeneously low (HL)}-players with different level of abilities, and the average abilities are low; {\em heterogeneously high (HH)}-players with  different level of abilities, and the average abilities are high; and {\em distinct (DI)}- a portion of players with low average abilities, the other portion with high average abilities. Each player's ability was generated according to a Gaussian distribution with means that follow the probability density function (See Figure~\ref{fig:pdfs}). Let $\rho_{v}$ be player $v$'s ability, then his cost coefficient $\delta_{v} = 1- \rho_{v}$.

In our experiments, we set $\lambda = 0.5$ as it was standard in many geometric reward mechanisms. In practice, a stakeholder usually sets $\varphi \leq 1$ to make profits, but  $\varphi$ should be as close to 1 as possible to encourage players to participate.  We let $\varphi = 1$. To encourage players to join, we set $\mu = 0.9$, and $\phi = \varphi - \mu = 0.1$.  Note that $\eta \geq \lambda / 2 = 0.25$, we set $\eta = 0.25$. 
For each group of players in each dataset, we varied the noise factors from 0 to 1 with an increment of 0.05. For each result (i.e., a data point) obtained, we ran the respective experiment 20 times. We ran all the experiments on the  same 3.7GHz 6-core Linux machine with 32GB RAM. 

\subsection{Results}
\begin{figure}
\centering
\begin{subfigure}[b]{.23\textwidth}
 \includegraphics[width=.9\columnwidth]{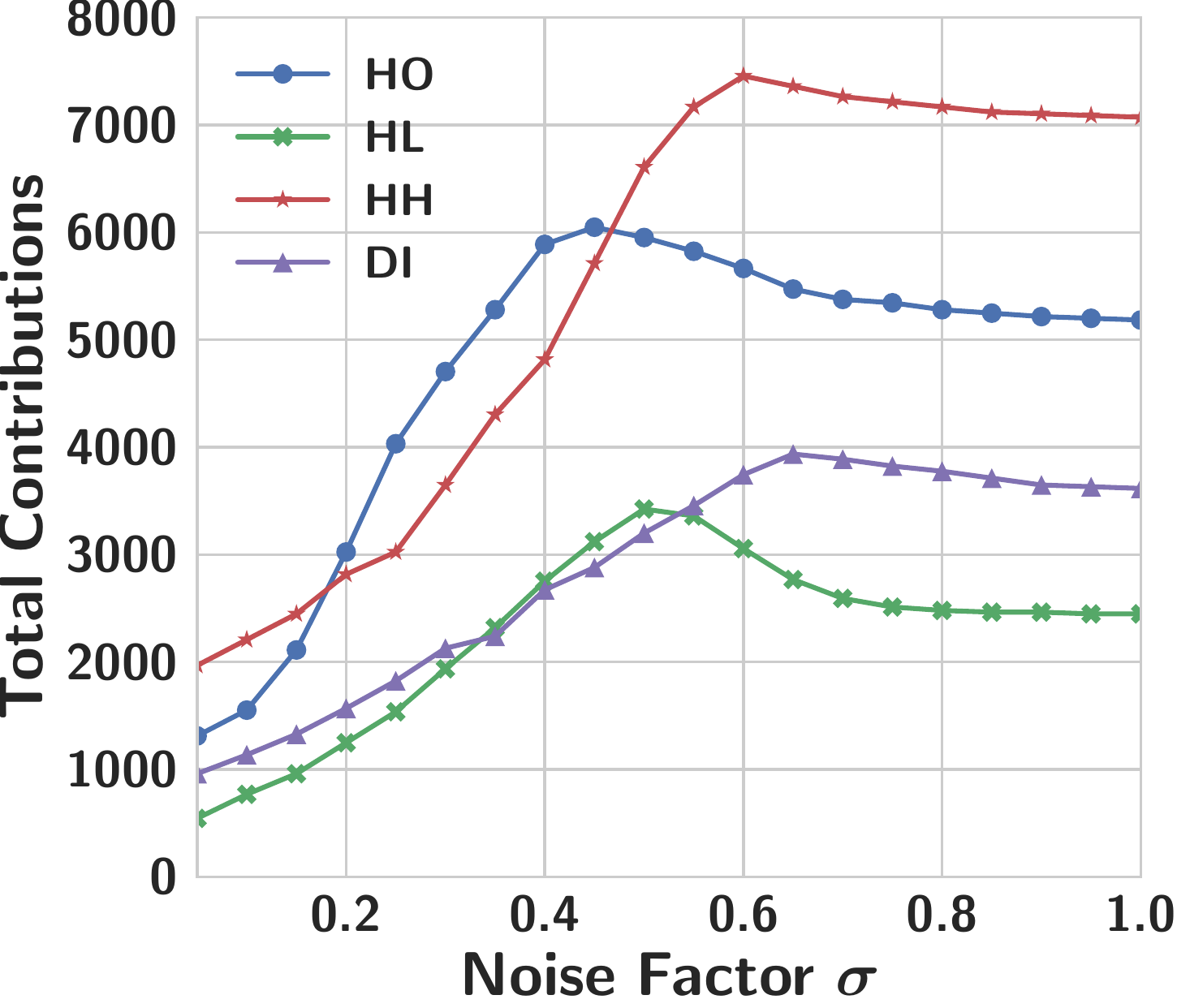}
\caption{Twitter.}
\label{fig:twitterbias}
\end{subfigure}
\hfill
\begin{subfigure}[b]{.23\textwidth}
 \includegraphics[width=.9\columnwidth]{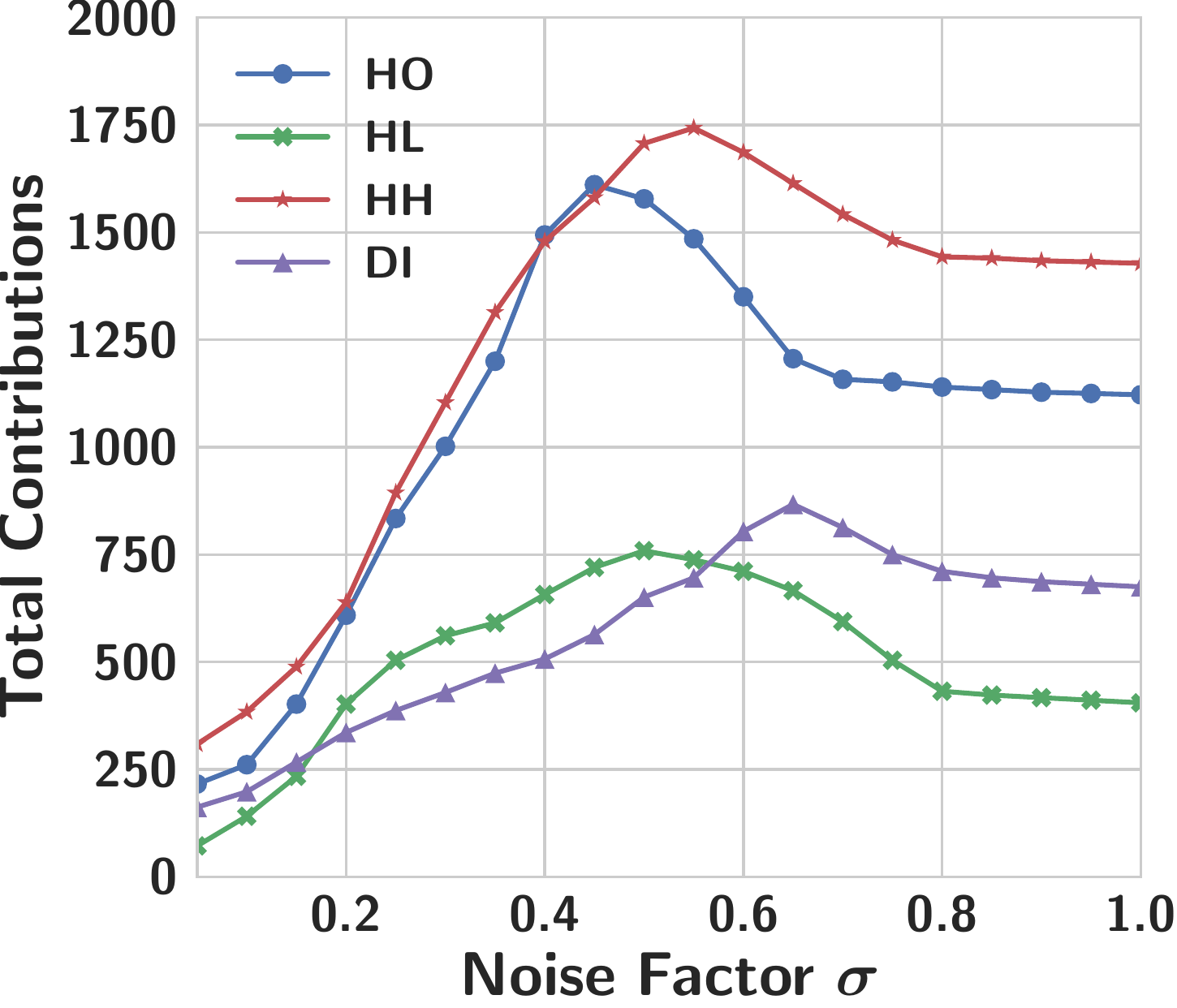}
\caption{Flickr.}
\label{fig:flickrbias}
\end{subfigure}
\hfill
\begin{subfigure}[b]{.23\textwidth}
 \includegraphics[width=.9\columnwidth]{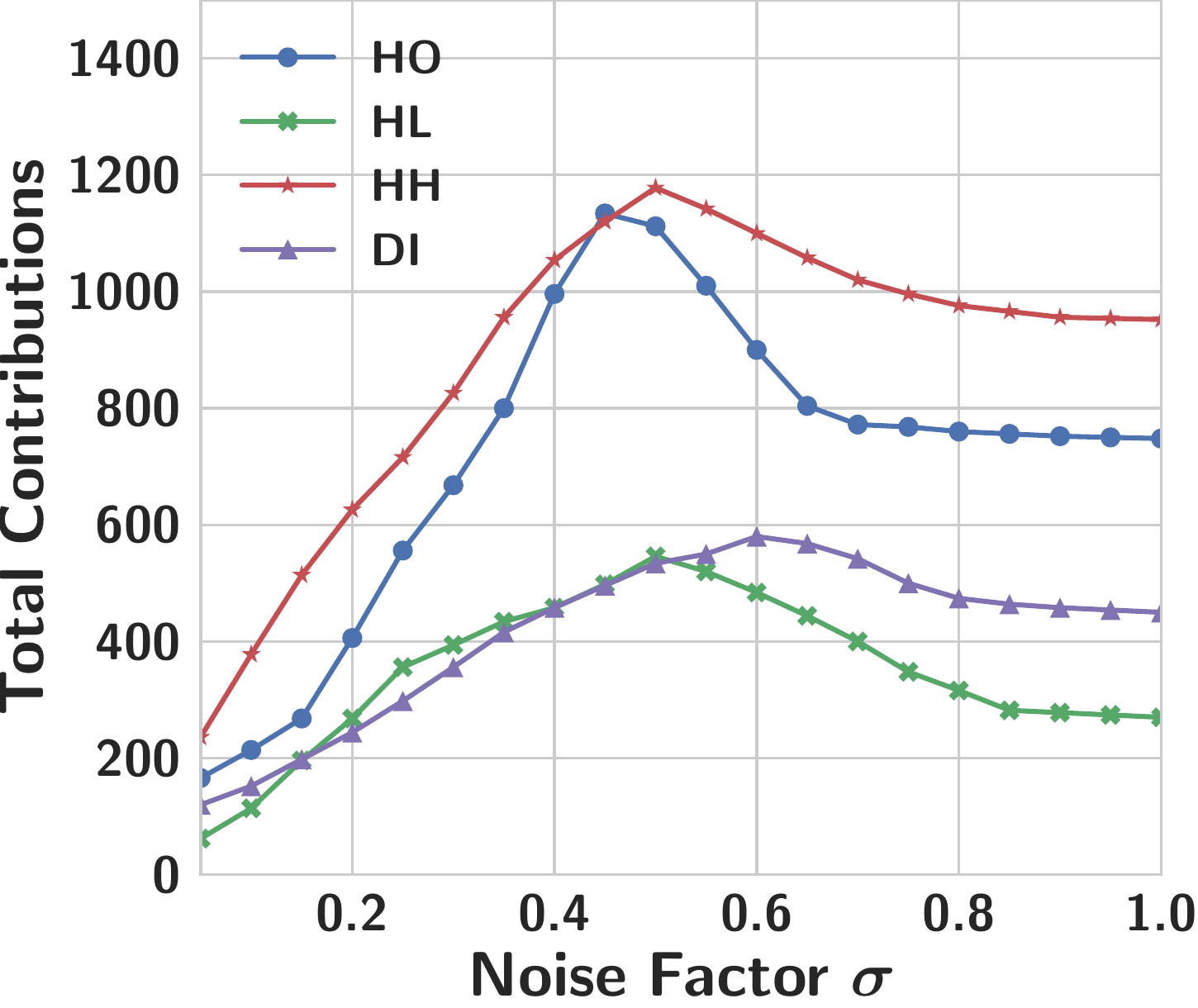}
\caption{Flixster.}
\label{fig:flixsterbias}
\end{subfigure}
\hfill
\begin{subfigure}[b]{.23\textwidth}
 \includegraphics[width=.9\columnwidth]{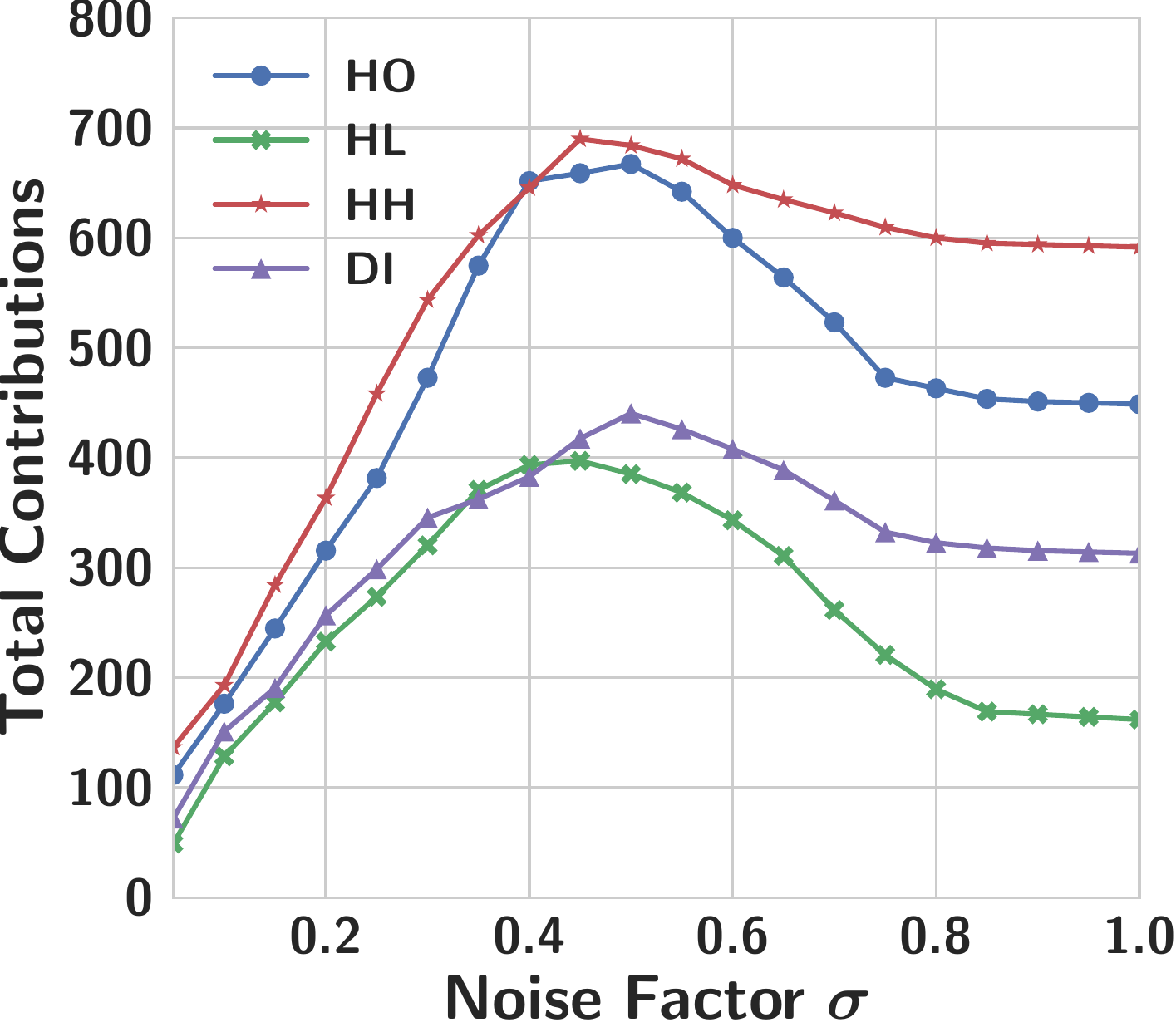}
\caption{Digg.}
\label{fig:diggbisa}
\end{subfigure}

\caption{A comparison of total contributions with different noise factors.}
\label{fig:totalcontribution}
\end{figure}

\begin{figure}[ht]
\centering
\begin{subfigure}[b]{.23\textwidth}
 \includegraphics[width=.9\columnwidth]{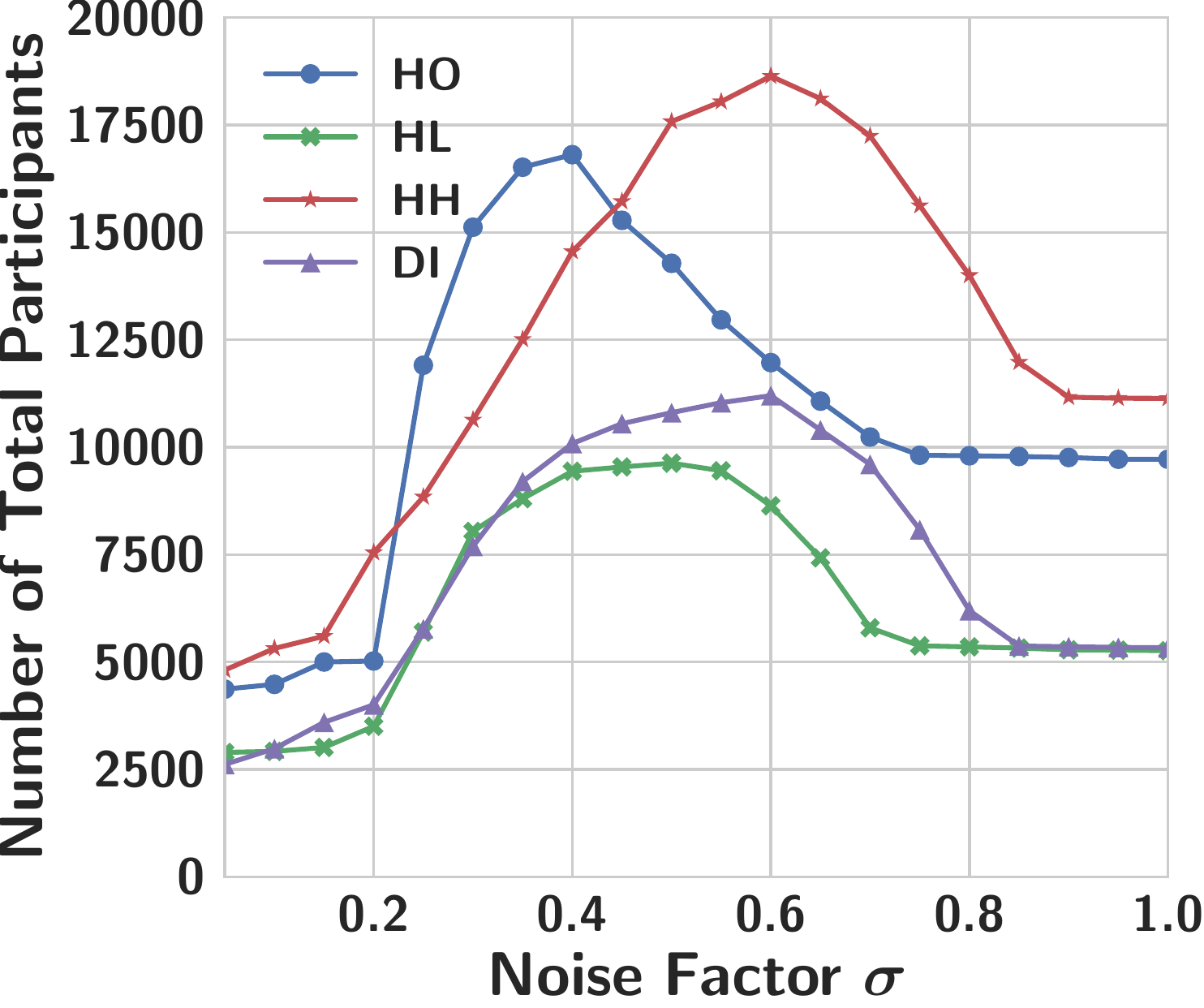}
\caption{Num. of Players.}
\label{fig:numofptwitter}
\end{subfigure}
\hfill
\begin{subfigure}[b]{.23\textwidth}
 \includegraphics[width=.9\columnwidth]{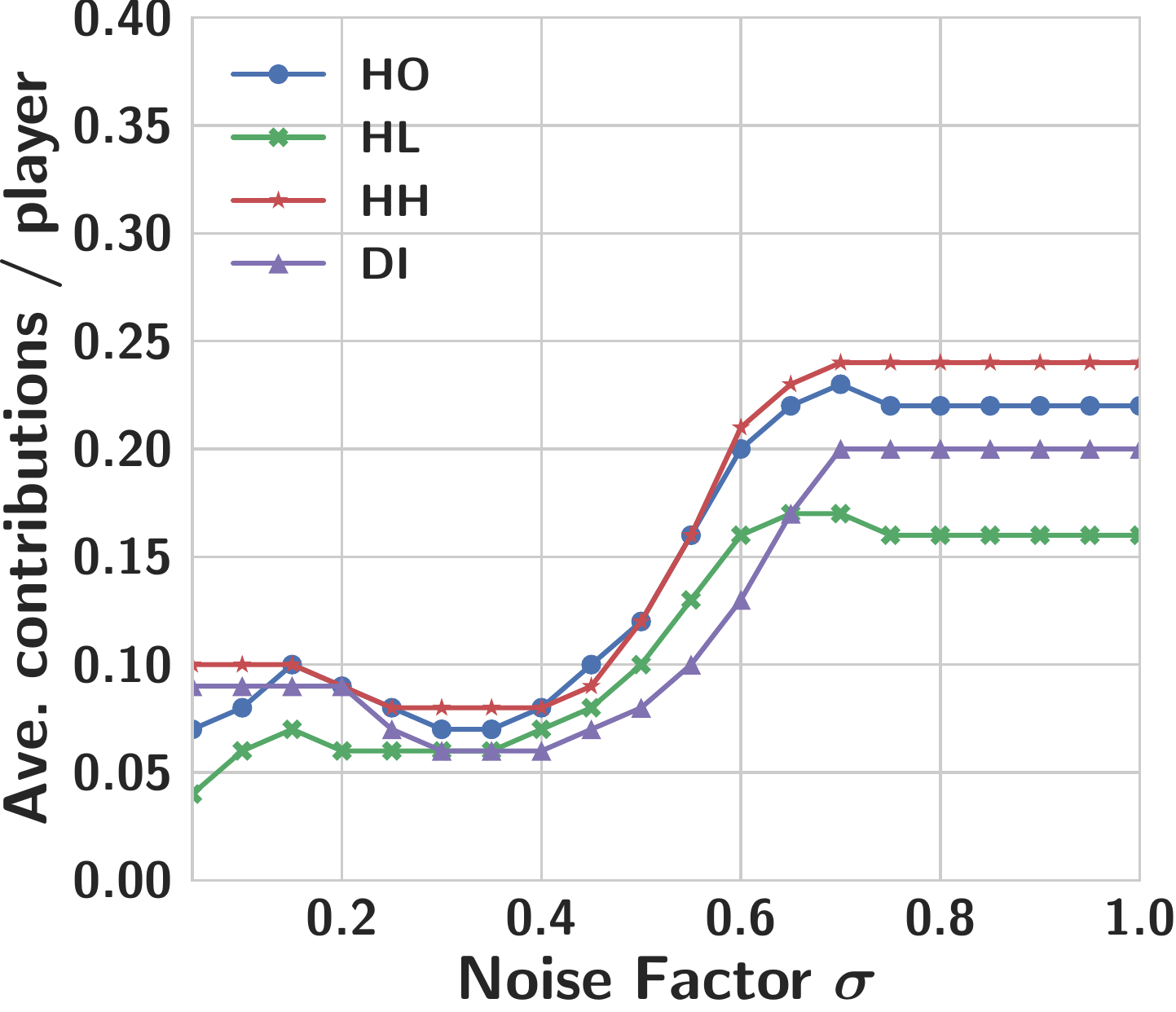}
\caption{Ave. Contributions.}
\label{fig:avecontwitter}
\end{subfigure}
\caption{A comparison of number of participants  and average contributions per player on Twitter.}
\label{fig:comparetwitter}
\end{figure}

For each dataset, we compared the total aggregated task efforts by each group of players as the noise factor $\sigma$ varied (See Figure~\ref{fig:totalcontribution}). Figure~\ref{fig:twitterbias} shows that  the total contributions were quite low ( $\leq 2,000$) when the noise factor was zero. The main reason is that players had no incentives to make extra diffusion contributions (See Figure~\ref{fig:numofptwitter})  since the diffusion rewards were determined by a random lottery with no dependence upon the efforts of players when $\sigma=0$. 

As $\sigma$ began to increase, the total contributions experienced a significant growth, which was largely due to the increase of players (See Figure~\ref{fig:numofptwitter}). The growth continued until it reached the peak where the total contributions began to shrink gradually (See Figure~\ref{fig:totalcontribution}). The decline continued until it started to level off when the noise factor exceeded 0.8 (See Figure~\ref{fig:totalcontribution}). An explanation for this trend is that when $\sigma$ increased, the MWC mechanism values higher diffusion efforts more. When $\sigma$ is low, the reward mechanism favors players with average abilities. These players usually take up a large portion of the population. When $\sigma$ reached a point, the mechanism no longer leaned toward these players. It became more appealing to high-ability players than the low-ability or ordinary-ability players (See Figure~\ref{fig:avecontwitter}). However, all groups except the HH group included a small portion of high-ability players.  That also explains why the HH group experienced a slighter decline than the other groups. Similar observations were found in social networks.  

Players in the HH and HO groups typically performed better than the other two groups.  As the noise factor $\sigma$ increased, the HO groups experienced the highest degree of growth in total task efforts.  Two factors contributed to this phenomenon. First, HO groups mainly consisted of players with normal abilities (See Figure~\ref{fig:hopdf}). Second, when $\sigma$ increased to a point, the contests favored players with common abilities.   Figure~\ref{fig:totalcontribution} shows that the total contributions of the DI groups were the most invariant to the noise factors. This is because the DI groups lacked a population mass of players with ordinary abilities (See Figure~\ref{fig:dipdf}). The observations applied to all the three datasets  (See Figure~\ref{fig:totalcontribution}). Despite that the optimal noise factors differed in  populations and network structures,  there were ``sweet spots" for stakeholders to maximize the total efforts. In our experiments, all the optimal noise factors fell within the range between 0.45 and 0.65 (See Figure~\ref{fig:totalcontribution}), which suggested that medium noise factors were typically superior than the others.

To study how false-name attacks affect players' rewards, we measured the total rewards earned by a player that had created different numbers of false identities. For each network,  we considered a population consisting of the same percentage ($25\%$) of players from each of  the four groups: HO, HL, HH,  and DI.  We compared four noise factors: 0.4, 0.5, 0.6, and 0.7 with all the four networks. 

\begin{figure}
\centering
\begin{subfigure}[b]{.23\textwidth}
 \includegraphics[width=.9\columnwidth]{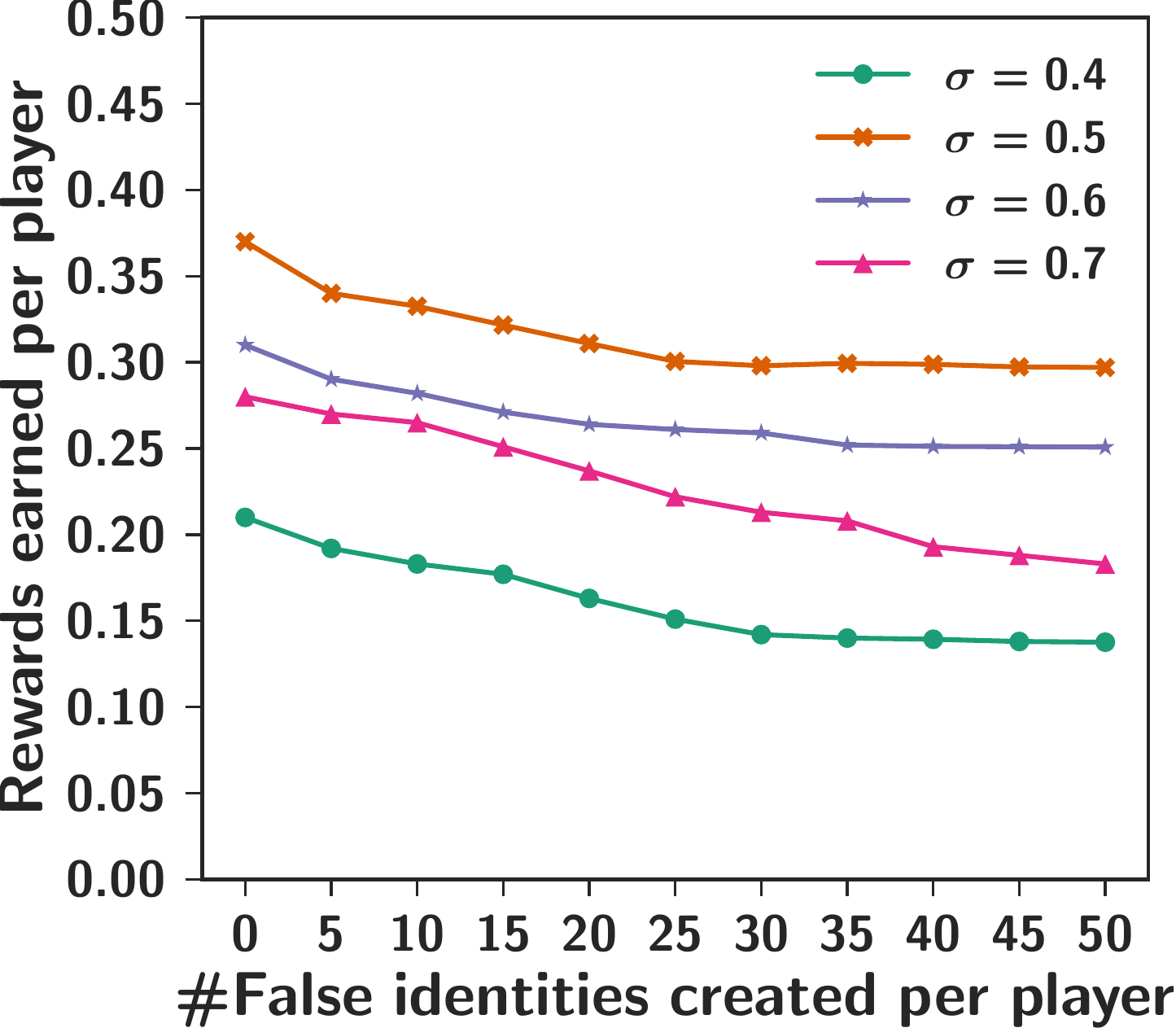}
\caption{Twitter.}
\label{fig:sybiltwitter}
\end{subfigure}
\hfill
\begin{subfigure}[b]{.23\textwidth}
 \includegraphics[width=.9\columnwidth]{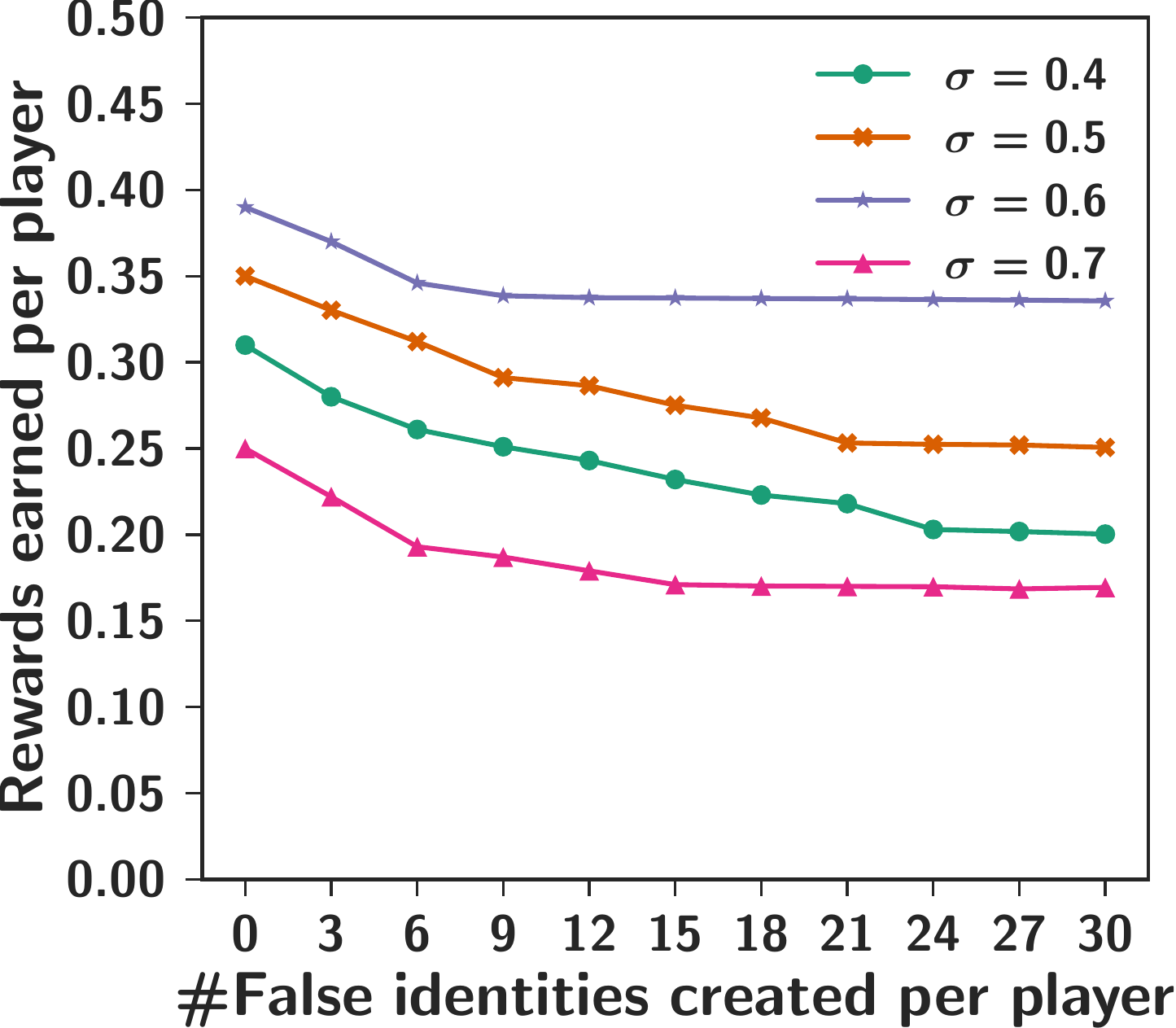}
\caption{Flickr.}
\label{fig:sybilflickr}
\end{subfigure}
\hfill
\begin{subfigure}[b]{.23\textwidth}
 \includegraphics[width=.9\columnwidth]{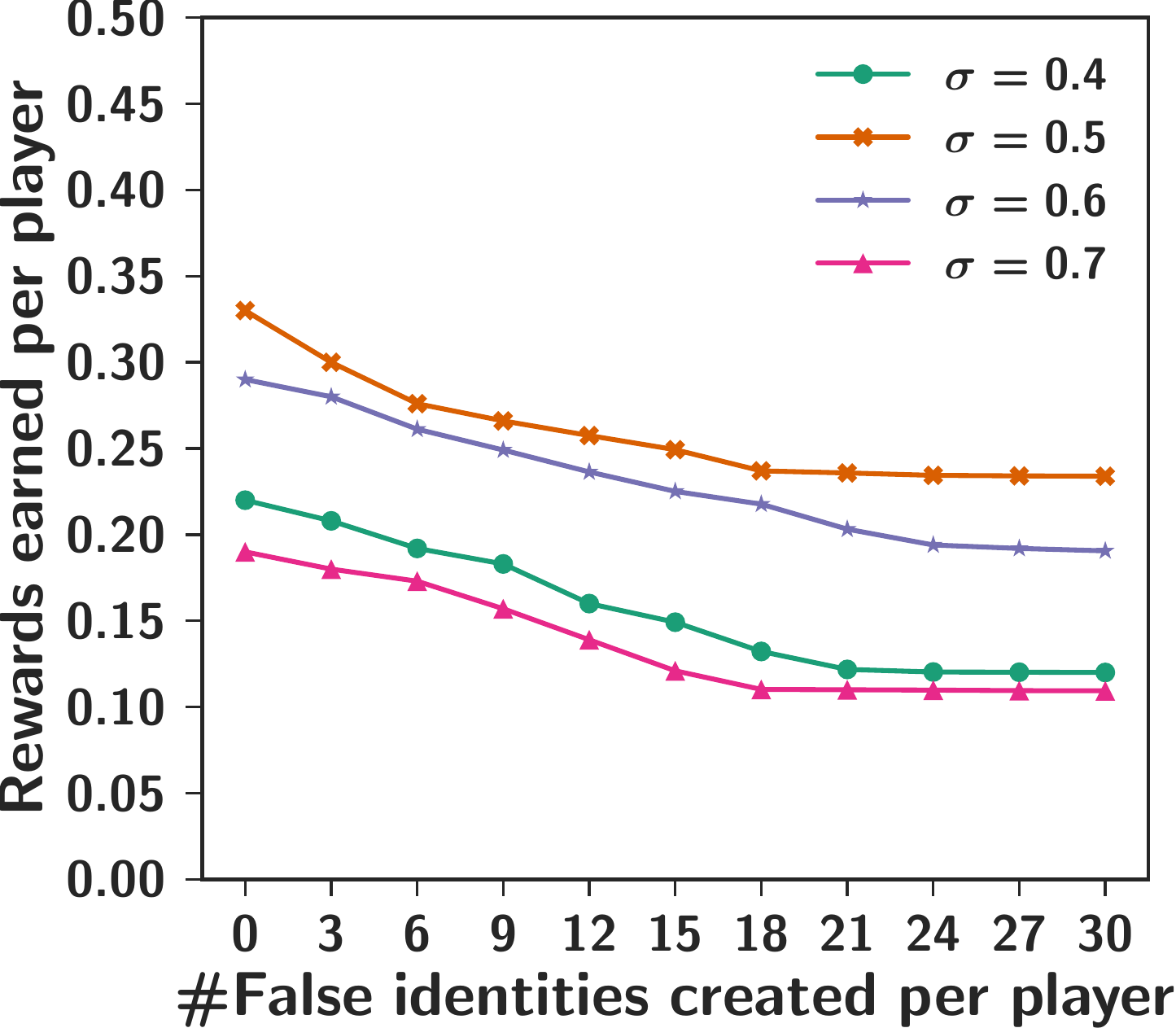}
\caption{Flixster.}
\label{fig:sybilflixster}
\end{subfigure}
\hfill
\begin{subfigure}[b]{.23\textwidth}
 \includegraphics[width=.9\columnwidth]{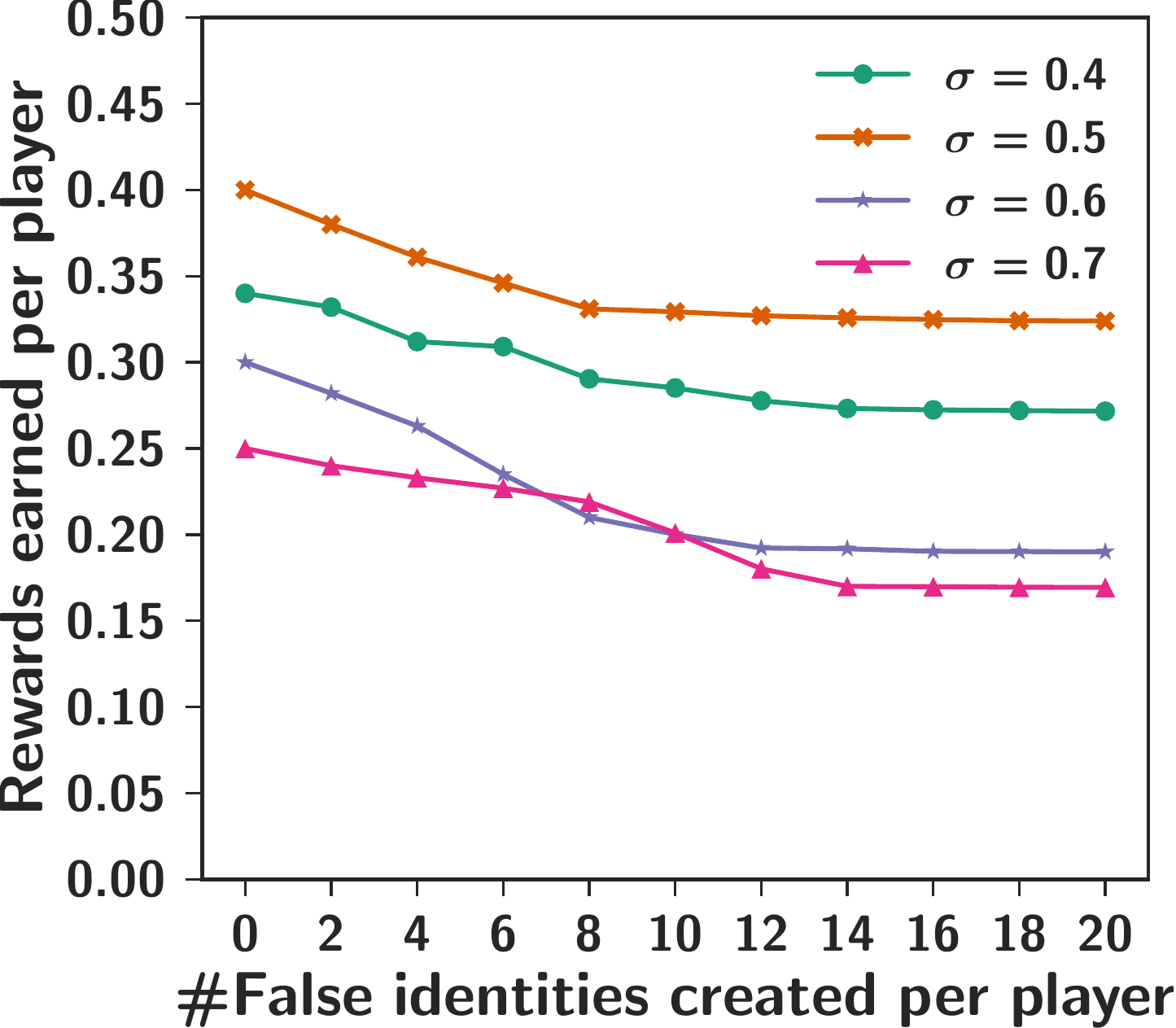}
\caption{Digg.}
\label{fig:sybildigg}
\end{subfigure}

\caption{Rewards (normalized) obtained by a player that creates different number of false identities for groups with different noise factors.}
\label{fig:rewardsybils}
\end{figure}

Results show that the rewards earned by each player reduced significantly as the number of false identities increased from zero. The decline then slowed down until the rewards remained almost steady (See Figure~\ref{fig:rewardsybils}).  These trends were observed on all the groups of experiments that were given different noise factors and different network structures.  An explanation is that when a player created a small number of false identities, his task efforts would be diluted to each false identity. This substantially reduced the player's virtual credits and  his diffusion rewards. When the number of false identities continued to increase, there was little room to reduce.  This consistent results further echoed our theoretical analysis that the MWC mechanism is false-name-proof.

In summary,  experimental results indicate that stakeholders can maximize the total task contributions by selecting appropriate  (typically moderate) noise factors for the contests. It further demonstrates that players will not gain by creating false identities. 
The results also show that the MWC mechanism can scale to large social networks with hundreds of thousands of nodes.
\section{Conclusion}
In this paper, we have introduced a novel multi-winner contests mechanism for strategic diffusion in social networks. The mechanism is false-name-proof and individual rational for players with successful referrals.   It is computationally efficient, budget-constrained, monotonic,  and subgraph-constrained.  Experiments on four real-world social networking datasets show that stakeholders can boost the performance of players by selecting proper noise factors. They further indicate that the MWC mechanism is both false-name-proof and scalable. Our work demonstrates the promising prospects of bringing contests to mechanism design.

Our work opens several exciting avenues for future research.  In our model, agents do not discount the future and do not have uncertainties about the delivery of the rewards. In many real-world activities (e.g., crowdfunding, and investments), however, agents need to consider future risks when they make decisions. Another fertile area is to develop novel methods to automatically select optimal noise factors of MWC mechanisms for different concerns (e.g., profit maximization).  We also find it very rewarding to integrate contests into the design of truthful mechanisms~\cite{shen2016online,zhao2018selling}.

\section{Appendix}
\subsection{Proof of Theorem 1}
\begin{proof}
By Definition~\ref{defn:falsenameattack},  the task rewards for player $v$ remain unchanged even if he creates replicas. To prove $\pi (v) \geq \sum_{r\in R} \pi (r)$, it suffices that if the $v$'s diffusion rewards are greater than or equal to the sum of diffusion reward of her false identities. That is, $\pi_{d}(v) \geq \sum_{r\in R} \pi_{d}(r)$. To prove this condition holds,  we first show that the virtual credits earned by $v$ are strictly greater than the sum of the virtual credits earned by the replicas $R$. That is, $b_{v} > \sum_{r\in R} b_{r}$. There are three scenarios according the types of false-name attacks: type 1 attacks, type 2 attacks, and hybrid  attacks. We prove each case in turn:\\
\begin{enumerate}
\item[$\bullet$] Type 1 attacks: By Equation~\ref{eq:virtualcredits}, the virtual credits of original node  $v$: $b_{v} = \eta \cdot (t_{v})^{2} + t_{v} \cdot \sum_{u \in \kappa^{+}_{v}} \sum_{p \in P_{vu}} t_{u} \cdot \omega(p) \cdot \lambda^{|p|}$.  For simplicity, we let $b(v) =  \eta \cdot (t_{v})^{2} + t_{v} \cdot C(v)$. 

Note that node $v$ shares the same successors with the replica $r_{m}$. Therefore, we have the virtual credits for replica $r_{m}$: $ b_{r_{m}} = \eta \cdot (t_{r_{m}})^{2}  + t_{r_{m}} \cdot \sum_{u \in \kappa^{+}_{r}} \sum_{p \in P_{ru}} t_{u} \cdot \omega(p) \cdot \lambda^{|p|}$ = $\eta \cdot (t_{r_{m}})^{2}  + t_{r_{m}} \cdot \sum_{u \in \kappa^{+}_{v}} \sum_{p \in P_{vu}} t_{u} \cdot \omega(p) \cdot \lambda^{|p|}$ = $\eta \cdot (t_{r_{m}})^{2} + t_{r_{m}}  \cdot C(v)$.\\

Similarly, we have  $r_{m-1}$: $b_{r_{m-1}} = \eta \cdot (t_{r_{m-1}})^{2} + t_{r_{m-1}} \sum_{u \in \kappa^{+}_{r_{m-1}}} \sum_{p \in P_{r_{m-1}u}} t_{u}\cdot \omega(p)\cdot \lambda^{|p|}$ = $\eta \cdot (t_{r_{m-1}})^{2} +  t_{r_{m-1}}  \cdot \sum_{u \in \kappa^{+}_{r_{m-1}}} \sum_{p \in P_{r_{m-1}u}} t_{u} \cdot \omega(p) \cdot \lambda^{|p|}$ = $\eta \cdot (t_{r_{m-1}})^{2} + \lambda t_{r_{m-1}}\cdot \sum_{u \in \kappa^{+}_{r_{m}}} \sum_{p \in P_{r_{m}u}} t_{u} \cdot \omega(p) \cdot \lambda^{|p|} + \lambda t_{r_{m-1}} \cdot t_{r_{m}} \eta \cdot (t_{r_{m-1}})^{2} + \lambda t_{r_{m-1}}\cdot  C(v) + \lambda \cdot  t_{r_{m-1}} \cdot t_{r_{m}} $.\\

Likewise, we have: $b_{r_{m-2}} = \eta \cdot (t_{r_{m-2}})^{2} + \lambda^{2} t_{r_{m-2}} \cdot C(v) + \lambda^{2}  t_{r_{m-2}} \cdot t_{r_{m}} + \lambda t_{r_{m-1}} t_{r_{m-2}} $, ..., $b_{v'_{1}} = \eta \cdot (t_{r_{1}})^{2} + \lambda^{m-1} t_{r_{1}} \cdot  C(v)  + \lambda t_{r_{1}}\cdot t_{r_{2}} + \lambda^{2}  \cdot t_{r_{1}}\cdot t_{r_{3}} + ... +  \lambda^{m-1} t_{r_{1}} \cdot t_{r_{m}}$.\\

Now we have the sum of the virtual credits for all replicas:
\begin{equation*}
\begin{split}
\sum_{r \in R} b_{r} &= b_{r_{1}} +  ... + b_{r_{m-1}} + b_{b_{m}}  \\
&= \eta \cdot [(t_{r_{1}})^2 + ... +  (t_{r_{m-1}})^{2} +(t_{r_{m}})^{2}  ] \\
& [\lambda^{m-1} t_{r_{1}} + \lambda^{m-2}  t_{r_{2}} + ... +\lambda t_{r_{m-1}}+ t_{r_{m}}] +  t_{r_{m}} \cdot \\
&  C(v) + \lambda t_{r_{1}}\cdot t_{r_{2}} + \lambda^{2}  \cdot t_{r_{1}}\cdot t_{r_{3}} + ... +  \lambda t_{r_{m-1}} \cdot t_{r_{m}}\\
&< \eta \cdot [(t_{r_{1}})^2 + ... +  (t_{r_{m-1}})^{2} +(t_{r_{m}})^{2}  ] + [ t_{r_{1}} +\\
& t_{r_{2}} + ... +t_{r_{m}}] \cdot C(v)+  t_{r_{1}}\cdot t_{r_{2}} +   ... +   t_{r_{m-1}} \cdot t_{r_{m}}\\
&= \eta [t_{r_{1}} +  t_{r_{2}} + ... +t_{r_{m}}]^{2} + t_{v}\cdot C(v) + \\
&(\lambda - 2\eta)[t_{r_{1}}\cdot t_{r_{2}} +   ... +   t_{r_{m-1}} \cdot t_{r_{m}}]\\
&\leq \eta (t_{v})^{2} + t_{v}\cdot C(v)\\
&= b(v) \;.
\end{split}
\end{equation*}
The first inequality holds since $0<\lambda <1$.  The second inequality holds because $\lambda \geq \eta/2$. Thus,  $b_{v} > \sum_{r \in S} b_{r}$. \\

\item[$\bullet$] Type 2 attacks:  Note that replicas $r_{1}, ..., r_{m}$ share the same successors with node $v$: 
\begin{equation*}
\begin{split}
 b_{r_{j}} &= \eta \cdot (t_{r_{j}})^{2}  + t_{r_{j}} \cdot \sum_{u \in \kappa^{+}_{r_{j}}} \sum_{p \in P_{r_{j}u}} t_{u} \cdot \omega(p) \cdot \lambda^{|u|}\\
 &= \eta \cdot (t_{r_{j}})^{2} + t_{r_{j}} \cdot \sum_{u \in \kappa^{+}_{v}}\sum_{p \in P_{vu}} t_{u} \cdot \omega(p) \cdot \lambda^{|p|}\\
  &= \eta \cdot (t_{r_{j}})^{2} + t_{r_{j}}  \cdot C(v) \;. 
  \end{split}
\end{equation*}
\\

Thus, we have:
\begin{equation*}
\begin{split}
\sum_{r \in R} b_{r} &= \eta \cdot [(t_{r_{1}})^2 + ... +  (t_{r_{m-1}})^{2} +(t_{r_{m}})^{2}  ] + \\
& [ t_{r_{1}} +  t_{r_{2}} + ... +t_{r_{m}}] \cdot C(v)\\
&< \eta \cdot [t_{r_{1}} +  t_{r_{2}} + ... +t_{r_{m}}]^{2} + t_{v} \cdot C(v) \\
&= b_{v} \;.
\end{split}
\end{equation*}
The inequality is by Multinomial Theorem. Thus,  $b_{v} > \sum_{r \in R} b_{r}$. \\
%The inequality is by Multinomial Theorem.\\
\item[$\bullet$] Hybrid attacks: Since hybrid attacks are combinations of type 1 attacks, and type 2 attacks, it is trivial to know that under hybrid attacks  $b_{v} > \sum_{r \in R} b_{r}$. \\
\end{enumerate}
To prove $\pi_{d}(v) \geq \sum_{r \in R} \pi_{d}(r)$, it suffices if the following inequality holds: 
\begin{equation}
\label{eq:sybilsinequation}
 \frac{(b_{v})^{\sigma}}{\sum_{u\in V_{G_{v}}} (b_{u})^{\sigma}} > \sum_{r \in R}   \frac{(b_{r})^{\sigma}}{\sum_{u \in V_{G_{r}}} (b_{u})^{\sigma}} \;.
\end{equation}
Note that $G_{r} = G_{v}\setminus\{v\} \cup R$.  Let $x = (b_{v})^{\sigma}$, $y=\sum_{u \in V_{G_{v}}} (b_{u})^{\sigma}$, $z=\sum_{r \in R} (b_{u})^{\sigma}$, and $q=\sum_{u\in V_{G_{r}}} (b_{u})^{\sigma}$, we have $0<\frac{x}{y}<1$,  $x > z>0$, and $y> q>0$.  Equation~\ref{eq:sybilsinequation} is equivalent to: $\frac{x}{y} > \frac{z}{q}$. Since $0<\frac{x}{y}<1$, we have:
%$\frac{z}{q}/ \frac{x}{y} = \frac{x-d}{y-d} /\frac{x}{y} = \frac{(x-d)y}{x(y-d)} =\frac{1-d/x}{1-d/y} < 1$. 
\begin{equation}
\label{eq:addingtrick}
\frac{z}{q}/ \frac{x}{y} = \frac{x-d}{y-d} /\frac{x}{y} = \frac{(x-d)y}{x(y-d)} =\frac{1-d/x}{1-d/y} < 1\; .
\end{equation}
Thus, $\frac{x}{y} > \frac{z}{q}$. Therefore, $\pi_{d}(v) \geq \sum_{r \in RS} \pi_{d}(r)$. Now it follows that the reward mechanism $\pi$ is false-name proof.
\end{proof}
\subsection{Proof of Theorem 2}
\begin{proof}
By Equations~\ref{eq:utility} and~\ref{eq:totalreward} , we have player $v$'s expected utility: $U(v) = \mu \cdot t_{v} + \pi_{d}(v) - \delta_{v}\cdot t_{v} = (\mu - \delta_{v})\cdot t_{v} + \pi_{d}(v) $.
%\begin{equation*}
%U(v) = \mu \cdot t_{v} + r_{d}(v) - \delta_{v}\cdot t_{v} = (\mu - \delta_{v})\cdot t_{v} + r_{d}(v) 
%\end{equation*}
Since the parameter $\mu$ is known to the player in advance, $\mu - \delta_{v} \geq 0$. Since $\pi_{d}(v) > 0$ if player has exerted diffusion contributions, we have $U(v) > 0$ for all players that have exerted both task efforts and diffusion contributions. 
\end{proof}
\noindent\textbf{Proof of Theorem 3}
\begin{proof}[Proof]
By Equation~\ref{eq:totalreward},  $\sum_{v\in G}\pi(v) =  \sum_{v\in G} (\mu \cdot t_{v} + \pi_{d}(v))$.
%\begin{equation}
%\sum_{v\in G}\pi(v) =  \sum_{v\in G} (\mu \cdot t_{v} + r_{d}(v))
%\end{equation}
By Equation~\ref{eq:diffreward},  $\sum_{v \in G} \pi_{d}(v) \leq \phi \sum_{v\in G} t_{v}$. 
%\begin{equation}
%\sum_{v \in G} r_{d}(v) \leq \phi \sum_{v\in G} t_{v}
%\end{equation}
Thus, we have $\sum_{v\in G}\pi(v) \leq (\mu + \phi) \sum_{v \in G}$.
%\begin{equation}
%\sum_{v\in G}\pi(v) \leq (\mu + \phi) \sum_{v \in G}
%\end{equation}
Let $\vartheta = \mu + \phi$, we have $\sum_{v\in G}\pi(v) \leq \vartheta \sum_{v \in G}$. 
\end{proof}
\subsection{Proof of Theorem 4}
\begin{proof}
Consider that the virtual credits of $v_{1}$ is $b_{v_{1}} = B$ when no successor is added. Now adding a direct successor $v_{i}$ to $v_{2}$, $v_{1}$'s virtual credits $b_{v_{1}^{'}} = B+ \sum_{p \in P_{v_{1}v{i}}}\lambda^{|p|}\cdot t_{v_{1}} \cdot t_{v_{i}} \cdot \omega(p)$. Similarly, if adding a direct successor of a successor of $v_{2}$, we have $v_{1}$'s virtual credits: $b_{v_{1}^{''}} = B+ \sum_{p \in P_{v_{1}v{j}}}\lambda^{|p|}\cdot t_{v_{1}} \cdot t_{v_{j}} \cdot \omega(p)$. Note that $dist(v_{1}, v_{i}) < dist(v_{1}, v_{j})$, and $t_{v_{i}} = t_{v_{j}}$, it suffices to show that $b'_{v_{1}} > b''_{v_{1}}$.  Following the same method shown in Equation~\ref{eq:addingtrick}, we have the diffusion rewards: $\pi_{d}(v_{1}^{'}) \geq  \pi_{d} (v_{1}^{''})$.
\end{proof}
\subsection{Proof of Theorem 5}
\begin{proof}
According to Equation~\ref{eq:diffreward}, player $v$ virtual credits and consequentially the diffusion rewards are determined by a contest among players in the graph rooted at $v$, $G_{v}$. It follows that the MWC mechanism satisfies the subgraph constraint.  
\end{proof}

\bibliographystyle{aaai}
\small
\bibliography{contests}

\begin{thebibliography}{}

\bibitem[\protect\citeauthoryear{Brill \bgroup et al\mbox.\egroup
  }{2016}]{brill2016false}
Brill, M.; Conitzer, V.; Freeman, R.; and Shah, N.
\newblock 2016.
\newblock False-name-proof recommendations in social networks.
\newblock In {\em AAMAS 2016},  332--340.

\bibitem[\protect\citeauthoryear{Burnap \bgroup et al\mbox.\egroup
  }{2013}]{burnap2013simulation}
Burnap, A.; Ren, Y.; Papalambros, P.~Y.; Gonzalez, R.; and Gerth, R.
\newblock 2013.
\newblock A simulation based estimation of crowd ability and its influence on
  crowdsourced evaluation of design concepts.
\newblock In {\em the ASME 2013 International Design Engineering Technical
  Conferences}.

\bibitem[\protect\citeauthoryear{Cason, Masters, and
  Sheremeta}{2010}]{cason2010entry}
Cason, T.~N.; Masters, W.~A.; and Sheremeta, R.~M.
\newblock 2010.
\newblock Entry into winner-take-all and proportional-prize contests: An
  experimental study.
\newblock {\em Journal of Public Economics} 94(9-10):604--611.

\bibitem[\protect\citeauthoryear{Cason, Masters, and
  Sheremeta}{2018}]{Cason2018winner}
Cason, T.~N.; Masters, W.; and Sheremeta, R.
\newblock 2018.
\newblock Winner-take-all and proportional-prize contests: theory and
  experimental results.
\newblock {\em Journal of Economic Behavior and Organization}.

\bibitem[\protect\citeauthoryear{Cha, Mislove, and
  Gummadi}{2009}]{cha2009measurement}
Cha, M.; Mislove, A.; and Gummadi, K.~P.
\newblock 2009.
\newblock A measurement-driven analysis of information propagation in the
  flickr social network.
\newblock In {\em WWW 2009},  721--730.

\bibitem[\protect\citeauthoryear{Chaffey}{2016}]{chaffey2016global}
Chaffey, D.
\newblock 2016.
\newblock Global social media research summary 2016.
\newblock {\em Smart Insights: Social Media Marketing}.

\bibitem[\protect\citeauthoryear{Conitzer \bgroup et al\mbox.\egroup
  }{2010}]{conitzer2010false}
Conitzer, V.; Immorlica, N.; Letchford, J.; Munagala, K.; and Wagman, L.
\newblock 2010.
\newblock False-name-proofness in social networks.
\newblock In {\em International Workshop on Internet and Network Economics},
  209--221.
\newblock Springer.

\bibitem[\protect\citeauthoryear{Dow \bgroup et al\mbox.\egroup
  }{2012}]{dow2012shepherding}
Dow, S.; Kulkarni, A.; Klemmer, S.; and Hartmann, B.
\newblock 2012.
\newblock Shepherding the crowd yields better work.
\newblock In {\em CSCW 2012},  1013--1022.

\bibitem[\protect\citeauthoryear{Drucker and
  Fleischer}{2012}]{drucker2012simpler}
Drucker, F.~A., and Fleischer, L.~K.
\newblock 2012.
\newblock Simpler sybil-proof mechanisms for multi-level marketing.
\newblock In {\em EC 2012},  441--458.

\bibitem[\protect\citeauthoryear{Emek \bgroup et al\mbox.\egroup
  }{2011}]{emek2011mechanisms}
Emek, Y.; Karidi, R.; Tennenholtz, M.; and Zohar, A.
\newblock 2011.
\newblock Mechanisms for multi-level marketing.
\newblock In {\em EC 2011},  209--218.
\newblock ACM.

\bibitem[\protect\citeauthoryear{Ferrara \bgroup et al\mbox.\egroup
  }{2016}]{ferrara2016rise}
Ferrara, E.; Varol, O.; Davis, C.; Menczer, F.; and Flammini, A.
\newblock 2016.
\newblock The rise of social bots.
\newblock {\em Communications of the ACM} 59(7):96--104.

\bibitem[\protect\citeauthoryear{Galeotti and
  Goyal}{2009}]{galeotti2009influencing}
Galeotti, A., and Goyal, S.
\newblock 2009.
\newblock Influencing the influencers: a theory of strategic diffusion.
\newblock {\em The RAND Journal of Economics} 40(3):509--532.

\bibitem[\protect\citeauthoryear{Gao \bgroup et al\mbox.\egroup
  }{2015}]{gao2015survey}
Gao, H.; Liu, C.~H.; Wang, W.; Zhao, J.~R.; Song, Z.; Su, X.; Crowcroft, J.;
  and Leung, K.~K.
\newblock 2015.
\newblock A survey of incentive mechanisms for participatory sensing.
\newblock {\em IEEE Communications Surveys and Tutorials} 17(2):918--943.

\bibitem[\protect\citeauthoryear{Goyal, Bonchi, and
  Lakshmanan}{2010}]{goyal2010learning}
Goyal, A.; Bonchi, F.; and Lakshmanan, L.~V.
\newblock 2010.
\newblock Learning influence probabilities in social networks.
\newblock In {\em WSDM 2010},  241--250.

\bibitem[\protect\citeauthoryear{Goyal, Bonchi, and
  Lakshmanan}{2011}]{goyal2011data}
Goyal, A.; Bonchi, F.; and Lakshmanan, L.~V.
\newblock 2011.
\newblock A data-based approach to social influence maximization.
\newblock {\em the VLDB Endowment} 5(1):73--84.

\bibitem[\protect\citeauthoryear{Hodas and Lerman}{2014}]{hodas2014simple}
Hodas, N.~O., and Lerman, K.
\newblock 2014.
\newblock The simple rules of social contagion.
\newblock {\em Scientific Reports} 4:4343.

\bibitem[\protect\citeauthoryear{Hogg and Lerman}{2012}]{hogg2012social}
Hogg, T., and Lerman, K.
\newblock 2012.
\newblock Social dynamics of digg.
\newblock {\em EPJ Data Science} 1(1):5.

\bibitem[\protect\citeauthoryear{Ipeirotis}{2010}]{ipeirotis2010analyzing}
Ipeirotis, P.~G.
\newblock 2010.
\newblock Analyzing the amazon mechanical turk marketplace.
\newblock {\em XRDS: Crossroads, The ACM Magazine for Students} 17(2):16--21.

\bibitem[\protect\citeauthoryear{Jackson and
  Yariv}{2011}]{jackson2011diffusion}
Jackson, M.~O., and Yariv, L.
\newblock 2011.
\newblock Diffusion, strategic interaction, and social structure.
\newblock In {\em Handbook of Social Economics}, volume~1. Elsevier.
\newblock  645--678.

\bibitem[\protect\citeauthoryear{Jia, Skaperdas, and
  Vaidya}{2013}]{jia2013contest}
Jia, H.; Skaperdas, S.; and Vaidya, S.
\newblock 2013.
\newblock Contest functions: Theoretical foundations and issues in estimation.
\newblock {\em Int. J. Ind. Organ.} 31(3):211--222.

\bibitem[\protect\citeauthoryear{Kempe, Kleinberg, and
  Tardos}{2003}]{kempe2003maximizing}
Kempe, D.; Kleinberg, J.; and Tardos, {\'E}.
\newblock 2003.
\newblock Maximizing the spread of influence through a social network.
\newblock In {\em KDD 2003},  137--146.

\bibitem[\protect\citeauthoryear{Leskovec, Adamic, and
  Huberman}{2007}]{leskovec2007dynamics}
Leskovec, J.; Adamic, L.~A.; and Huberman, B.~A.
\newblock 2007.
\newblock The dynamics of viral marketing.
\newblock {\em ACM Transactions on the Web (TWEB)} 1(1):5.

\bibitem[\protect\citeauthoryear{Lorenz \bgroup et al\mbox.\egroup
  }{2011}]{lorenz2011social}
Lorenz, J.; Rauhut, H.; Schweitzer, F.; and Helbing, D.
\newblock 2011.
\newblock How social influence can undermine the wisdom of crowd effect.
\newblock {\em Proc. Natl. Acad. Sci. USA} 108(22):9020--9025.

\bibitem[\protect\citeauthoryear{Naroditskiy \bgroup et al\mbox.\egroup
  }{2014a}]{naroditskiy2014crowdsourcing}
Naroditskiy, V.; Jennings, N.~R.; Van~Hentenryck, P.; and Cebrian, M.
\newblock 2014a.
\newblock Crowdsourcing contest dilemma.
\newblock {\em J R Soc Interface} 11(99):20140532.

\bibitem[\protect\citeauthoryear{Naroditskiy \bgroup et al\mbox.\egroup
  }{2014b}]{naroditskiy2014referral}
Naroditskiy, V.; Stein, S.; Tonin, M.; Tran-Thanh, L.; Vlassopoulos, M.; and
  Jennings, N.~R.
\newblock 2014b.
\newblock Referral incentives in crowdfunding.
\newblock In {\em HCOMP 2014},  171--183.

\bibitem[\protect\citeauthoryear{Pickard \bgroup et al\mbox.\egroup
  }{2011}]{pickard2011time}
Pickard, G.; Pan, W.; Rahwan, I.; Cebrian, M.; Crane, R.; Madan, A.; and
  Pentland, A.
\newblock 2011.
\newblock Time-critical social mobilization.
\newblock {\em Science} 334(6055):509--512.

\bibitem[\protect\citeauthoryear{Rahwan \bgroup et al\mbox.\egroup
  }{2013}]{rahwan2013global}
Rahwan, I.; Dsouza, S.; Rutherford, A.; Naroditskiy, V.; McInerney, J.;
  Venanzi, M.; Jennings, N.~R.; and Cebrian, M.
\newblock 2013.
\newblock Global manhunt pushes the limits of social mobilization.
\newblock {\em Computer} 46(4):68--75.

\bibitem[\protect\citeauthoryear{Shen \bgroup et al\mbox.\egroup
  }{2018}]{shen2018information}
Shen, W.; Crandall, J.~W.; Yan, K.; and Lopes, C.~V.
\newblock 2018.
\newblock Information design in crowdfunding under thresholding policies.
\newblock In {\em AAMAS 2018},  632--640.

\bibitem[\protect\citeauthoryear{Shen, Lopes, and
  Crandall}{2016}]{shen2016online}
Shen, W.; Lopes, C.~V.; and Crandall, J.~W.
\newblock 2016.
\newblock An online mechanism for ridesharing in autonomous mobility-on-demand
  systems.
\newblock In {\em IJCAI 2016},  475--481.

\bibitem[\protect\citeauthoryear{Sheremeta}{2011}]{sheremeta2011contest}
Sheremeta, R.~M.
\newblock 2011.
\newblock Contest design: An experimental investigation.
\newblock {\em Economic Inquiry} 49(2):573--590.

\bibitem[\protect\citeauthoryear{Skaperdas}{1996}]{skaperdas1996contest}
Skaperdas, S.
\newblock 1996.
\newblock Contest success functions.
\newblock {\em Economic Theory} 7(2):283--290.

\bibitem[\protect\citeauthoryear{Tang \bgroup et al\mbox.\egroup
  }{2011}]{tang2011reflecting}
Tang, J.~C.; Cebrian, M.; Giacobe, N.~A.; Kim, H.-W.; Kim, T.; and Wickert,
  D.~B.
\newblock 2011.
\newblock Reflecting on the darpa red balloon challenge.
\newblock {\em Communications of the ACM} 54(4):78--85.

\bibitem[\protect\citeauthoryear{Todo, Iwasaki, and
  Yokoo}{2011}]{todo2011false}
Todo, T.; Iwasaki, A.; and Yokoo, M.
\newblock 2011.
\newblock False-name-proof mechanism design without money.
\newblock In {\em AAMAS 2011},  651--658.

\bibitem[\protect\citeauthoryear{Zhao \bgroup et al\mbox.\egroup
  }{2018}]{zhao2018selling}
Zhao, D.; Li, B.; Xu, J.; Hao, D.; and Jennings, N.~R.
\newblock 2018.
\newblock Selling multiple items via social networks.
\newblock In {\em AAMAS 2018},  68--76.

\end{thebibliography}
\end{document}